\RequirePackage{fix-cm}
\documentclass[twocolumn,epjc3]{svjour3}  

\smartqed  
\usepackage{color}
\usepackage{xcolor}
\definecolor{bred}{rgb}{0.8, 0.0, 0.0}
\definecolor{pblue}{rgb}{0.2, 0.2, 0.6}
\definecolor{ao}{rgb}{0.0, 0.5, 0.0}
\definecolor{carmine}{rgb}{0.59, 0.0, 0.09}

\newcommand{\CEvNS}{CE$\nu$NS}
\newcommand{\conusplus}{CONUS\texttt{+}}

\newcommand{\inches}{\ensuremath{{}^{\prime\prime}}}

\newcommand\Tstrut{\rule{0pt}{2.5ex}}         
\newcommand\Bstrut{\rule[-1.4ex]{0pt}{0pt}}   

\usepackage{graphicx}
\usepackage{amsmath}
\usepackage{amssymb}
\usepackage{hyperref}
\usepackage{cite}
\usepackage{upgreek}
\usepackage{subcaption}
\usepackage{url}
\usepackage{gensymb}
\usepackage{tabularx}
\usepackage{adjustbox}
\usepackage{arydshln}

\usepackage{hyperref}
\hypersetup{
    colorlinks = true,
    linkbordercolor = {white},
    linkcolor = bred,
    citecolor = pblue,
    urlcolor = pblue
}

\usepackage[switch]{lineno} 

\begin{document}\sloppy

\title{Background characterization of the \conusplus~experimental location }

\author{E.~S\'{a}nchez Garc\'{i}a\thanksref{e1, MPIK}, N.~Ackermann\thanksref{MPIK}, S.~Armbruster\thanksref{MPIK}, H.~Bonet\thanksref{MPIK}, C.~Buck\thanksref{MPIK}, K.~F\"{u}lber\thanksref{KBR}, J.~Hakenm\"{u}ller\thanksref{MPIK, Duke}, J.~Hempfling\thanksref{MPIK},  G.~Heusser\thanksref{MPIK}, E.~Hohmann\thanksref{PSI}, M.~Lindner\thanksref{MPIK}, W.~Maneschg\thanksref{MPIK},  K.~Ni\thanksref{MPIK}, M.~Rank\thanksref{KKL}, T.~Rink\thanksref{MPIK}, I.~Stalder\thanksref{KKL}, H.~Strecker\thanksref{MPIK}, R.~Wink\thanksref{KBR}, J.~Woenckhaus\thanksref{PSI}
}

\authorrunning{E.~S\'{a}nchez Garc\'{i}a et al.}
\institute{Max-Planck-Institut f\"ur Kernphysik, Saupfercheckweg 1, 69117 Heidelberg, Germany \label{MPIK}  \and PreussenElektra GmbH, Kernkraftwerk Brokdorf, Osterende, 25576 Brokdorf, Germany  \label{KBR} \and Kernkraftwerk Leibstadt AG, 5325 Leibstadt, Switzerland  \label{KKL} \and  Paul Scherrer Institut, Forschungsstrasse 111, 5232 Villigen, Switzerland\label{PSI} \and \emph{Present Address:} Duke University, NC 27708, USA\label{Duke} \vspace*{0.2cm} 
}
\thankstext{e1}{\href{mailto:esanchez@mpi-hd.mpg.de}{esanchez@mpi-hd.mpg.de (corresponding author)}}
\thankstext{e2}{\href{mailto:conus.eb@mpi-hd.mpg.de}{conus.eb@mpi-hd.mpg.de}}

\date{\today}

\maketitle

\begin{abstract}

\conusplus~is an experiment aiming at detecting coherent elastic neutrino-nucleus scattering (\CEvNS) of reactor antineutrinos on germanium nuclei in the fully coherent regime, continuing the CONUS physics program conducted at the Brokdorf nuclear power plant (KBR), Germany. The \conusplus~experiment is installed in the Leibstadt nuclear power plant (KKL), Switzerland, at a distance of 20.7~m from the 3.6~GW reactor core, where the antineutrino flux is $1.5\cdot 10^{13}$~s$^{-1}$cm$^{-2}$. The \CEvNS~signature will be measured with four point-contact high-purity low energy threshold germanium (HPGe) detectors. 

A good understanding of the background is crucial, especially events correlated with the reactor thermal power are troublesome, as they can mimic the predicted  \CEvNS~interactions. A large background characterization campaign  was conducted during reactor on and off times to find the best location for the \conusplus~setup. On-site measurements revealed a correlated, highly thermalized neutron field with a maximum fluence rate of   $(2.3\pm0.1)\cdot 10^{4}$~neutrons~d$^{-1}$cm$^{-2}$ during reactor operation. The $\gamma$-ray background was studied with a HPGe detector without shield, paying special attention to the thermal power correlated $^{16}$N decay and other neutron capture $\gamma$-lines. The muon flux was examined using a liquid scintillator detector measuring (107$\pm$3)~muons~s$^{-1}$m$^{-2}$, which corresponds to an average overburden of 7.4~m of water equivalent. 

The new background conditions in \conusplus~are compared to the previous CONUS ones, showing a 30~times higher flux of neutrons, but a 26~times lower component of reactor thermal power correlated $\gamma$-rays over 2.7~MeV. The lower \conusplus~overburden increases the number of muon-induced neutrons by 2.3~times and the flux of cosmogenic neutrons.  Finally, all the measured rates are discussed in the context of the \conusplus~background, together with the \conusplus~modifications performed to reduce the impact of the new background conditions at KKL. 


\end{abstract}

\section{Introduction}
\label{intro}

Coherent elastic neutrino-nucleus scattering (\CEvNS) is a fundamental interaction within the framework of the Standard Model (SM), first anticipated in 1974~\cite{Freedman:1973yd, Kopeliovich:1974mv}. It was measured for the first time by the COHERENT collaboration at a spallation neutron source (SNS), where neutrinos are generated through pion decays at rest~\cite{Coherent:2017,COHERENT:2020iec,Adamski:2024yqt}. Recently first indications of detection of $^{8}$B neutrinos from the Sun via \CEvNS~where reported by the XENON~\cite{XENON:2024ijk} and PandaX~\cite{PandaX:2024muv} collaborations. An alternative source of neutrinos for \CEvNS~investigations are nuclear reactors. Currently, there is a large effort worldwide to detect \CEvNS~at reactors~\cite{Aguilar-Arevalo:2024dln,Ackermann:2024kxo,MINER:2016igy,Colaresi:2022obx,NEON:2022hbk,NUCLEUS:2022zti,nGeN:2022uje,Akimov:2024lnl,Yang:2024exl,RELICS:2024opj,Ricochet:2023nvt,TEXONO:2024vfk}.

The \conusplus~project is a follow-up experiment of CONUS~\cite{Ackermann:2024kxo}, dedicated to detect \CEvNS~of reactor antineutrinos on germanium (Ge) nuclei in the fully coherent regime. CONUS was performed at the Brokdorf nuclear power plant (Kernkraftwerk Brokdorf; KBR) from 2018 to 2022, providing the best limit for \CEvNS~detection from a reactor to date, only a factor 1.6 over the prediction of the Standard Model~\cite{Ackermann:2024kxo}. However, the KBR reactor finished operation by the end of 2021. A new location was found at the Leibstadt nuclear power plant (Kernkraftwerk Leibstadt AG; KKL) in Switzerland for the successor \conusplus~experiment. The antineutrino flux will be measured by four p-type point-contact (PPC) high purity germanium (HPGe) detectors with an ultra low energy threshold around 160~eV$_{ee}$ (electron equivalent energy). The detectors are enclosed in a compact shield consisting of different layers of lead and partly borated polyethylene to reduce the environmental background. Additionally, two layers of scintillation plates equipped with photomultiplier tubes are installed as an active muon anti-coincidence system. More details about the detector design are provided in~\cite{CONUS:2024lnu}. 

A detailed understanding of all background contributions in the new experimental location is very important, particularly events correlated with the reactor thermal power can mimic \CEvNS~interactions. For example, escaping neutrons produced by the reactor nuclear fissions can induce nuclear recoils comparable to \CEvNS. For this reason, an extensive background characterization campaign was performed at the KKL power plant previous to the \conusplus~installation similar to the one conducted for CONUS at KBR. The neutron fluence was studied with thermal neutron counters and the neutron energy spectrum was directly determined with a Bonner Sphere array system~\cite{THOMAS200212} during reactor on and off times. Intense neutron fields can also produce high energy $\gamma$-rays by neutron capture reactions with different isotopes, that can lead to non stable background conditions in the detectors. In particular, $^{16}$N which is formed in (n,p) reactions on $^{16}$O in the water of the reactor pressure vessel, can contribute with a strong $\gamma$-ray emission up to 10~MeV at nuclear reactors~\cite{Hakenmuller:2019ecb}. At the same time, by detecting  $\gamma$-rays from neutron captures in materials in the vicinity of the detector, information on the  neutron fluence can be determined indirectly. The $\gamma$-ray background  up to 11~MeV was determined with a HPGe detector, that was previously used in the KBR power plant~\cite{Hakenmuller:2019ecb} for the same purpose. This allows for a direct comparison of the background data collected in both locations.

Background components that are not correlated with the reactor thermal power are also  relevant for the experiment. In particular, as the \conusplus~detector is located above ground, a large cosmogenic flux is expected. Galactic cosmic rays consist of high-energy particles, mostly protons, that upon impact with Earth's atmosphere, produce showers of secondary particles, some of which can reach the Earth's surface. This hadronic component can activate the Ge and Cu parts of the HPGe detectors, leading to the production of radioisotopes with short and long decay half-lifes. Additionally, muon-induced neutrons are created in the structural materials of the  reactor building as well as in the 
\conusplus~shield. At sea level, muons account for the majority of particles of the secondary cosmic rays with an expected flux of 189~muons~s$^{-1}$m$^{-2}$~\cite{Bogdanova:2006ex} and a mean energy of 4~GeV~\cite{ParticleDataGroup:2020ssz}. To significantly reduce the muon flux, a minimum overburden of a few 100's~m w.e. would be required. However, since nuclear power plants are typically located at surface, this becomes a critical background for \CEvNS~experiments at nuclear reactors~\cite{Bonet:2021wjw}. For this reason, the muon flux in the \conusplus~experimental location was determined using a liquid scintillator detector. Additionally, the suppression of the cosmogenic neutron flux in the \conusplus~location was thoroughly studied with detailed simulations.  


Another important source of background in the experiment is the air-borne inert gas radon (Rn) from natural radioactivity. High Rn concentration levels are expected in enclosed spaces that are surrounded by thick concrete walls and that lack from a proper fresh air ventilation system like in a reactor safety containment. Specifically, the $^{222}$Rn isotope with its half-life of 3.8~days can diffuse into the \conusplus~shield and decay inside the Ge detector chamber, increasing on this way the background level. In addition, the Rn concentration level has significant temporal fluctuations that are related to temperature variations within the building and to alterations in the air ventilation system, particularly during reactor outages. Additionally, during reactor safety and maintenance operations other inert gas radioisotopes such as $^{135}$Xe and $^{85}$Kr are released inside the reactor steel containment. 

Moreover, due to erosion/corrosion and during reactor maintenance operations artificially produced radioisotopes can be released,  that can be transported on dust particles and deposit on surfaces. A cross-contamination of the ultra-low background parts of a \conusplus-like detector could easily occur, enhancing the overall background. For this reason, the surface contamination in the \conusplus~room was carefully studied through several wipe test campaigns, in order to define specific detector handling and cleaning protocols to reduce the risk of a cross-contamination. 

This article reports the different background measurements performed for the \conusplus~experiment. 
It is organized as follows. The description of the experimental location and the environmental conditions is detailed in Sec.\,\ref{sec:1}. The measurement of the $\gamma$-ray and surface contamination background is described in Sec.\,\ref{sec:2}. The muon flux reduction and the overburden estimation is presented in Sec.\,\ref{sec:3}. The neutron energy spectrum measurements and simulations are discussed in Sec.\,\ref{sec:4}. Finally, the background conditions and compositions measured in both locations at KBR and KKL are compared in the conclusions\,\ref{conclusions}.

\section{The \conusplus~experimental location}
\label{sec:1}

\begin{figure}
    \centering
    \includegraphics[width=0.47\textwidth]{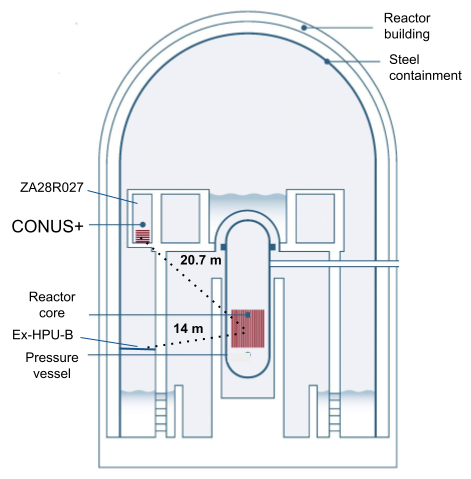}
    \caption{Scheme of the \conusplus~location inside the KKL containment~\cite{CONUS:2024lnu}. The ZA28R027 and the Ex-HPU-B locations evaluated during the background measurement campaign are also indicated.}
    \label{power_plant}
\end{figure}
\begin{figure}
    \centering
    \includegraphics[width=0.48\textwidth]{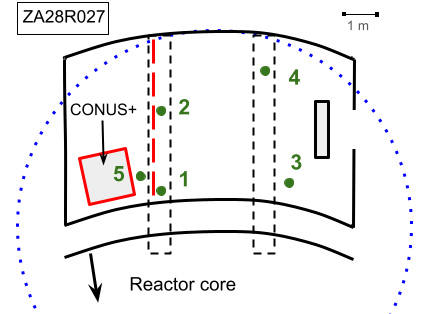}
    \caption{Scaled top view ZA28R027 room with the \conusplus~detector (red) installed. The experiment is isolated through a metallic wall (red dashed line). Two concrete reinforcements increased the roof thickness by 0.7~m (black dash lines). The five measurement positions studied during the background characterization campaign are indicated by green dots. The reactor drywell head with a diamter of 10~m and thickness of 3.8~cm is placed over the room in reactor off time (blue dashed line). }
    \label{conusplus_room}
\end{figure}

The \conusplus~experiment is installed in a dedicated room inside the reactor building of the KKL boiling water reactor (BWR). This room, named ZA28R027 from now on, is located 25.2~m above the ground surface, as shown in figure~\ref{power_plant}. This leads to a distance of 20.7~m between the center of the \conusplus~setup and the center of the reactor core which has a maximum thermal power of 3.6~GW. A second position at KKL named in the following Ex-HPU-B and used for auxiliary measurements within this work, is also indicated in figure~\ref{power_plant}. It is 14~m distant from the reactor core and accommodates a stainless steel platform of 5~m$^{2}$ on a metal grid, placed in an opened area at 8.6~m above ground. The position of the \conusplus~setup in ZA28R027 is depicted in figure~\ref{conusplus_room}. Although some permanent equipment from KKL is placed also in this room, an additional thin steel wall (indicated by a dashed red line) provides an isolated compartment for \conusplus. Five different positions named hereafter with 1 to 5 are marked with green dots; at these positions different measurements were conducted during the here presented background characterization campaign. The closest positions 1, 3 and 5 are at 20~m, 20.1~m and 20.2~m from the reactor core, while positions 2 and 4 are at 21.5~m and 22.2~m, respectively.    

The reactor building consists of a steel reinforced 1.2~m thick concrete dome and an inner 3.8~cm thick steel containment structure. The concrete roof of the ZA28R027 room has a thickness of 0.35~m and two 0.7~m thick crossbeams, as indicated by dashed black rectangles in figure~\ref{conusplus_room}. All these components provide an average overburden between 7-8~m water equivalent (m w.e.). This strongly suppresses the hadronic component of cosmic radiation. During the reactor maintenance periods the 10~m diameter reactor drywell head with a thickness of 3.8~cm is positioned over the room indicated by the dashed blue circle, increasing the overburden by approximately 0.25~m~w.e. More details about the overburden determination for \conusplus~are presented in Sec.\,\ref{sec:3}. 

The environmental conditions in the room (humidity, pressure, temperature and radon concentration) were monitored with a commercial RadonScout Home device from SARAD before the \conusplus~installation. The relative humidity was stable within an average of ($50\pm2$)\%  and the pressure varied within [968, 984]~mbar. The room temperature showed a mean value of 29$\degree$C and variations up to 4$\degree$C. Considering the maximum 5.6~kW consumption power of the full \conusplus~electronics setup and the reduced ventilation conditions in ZA28R027, the temperature in the room would raise above 35$\degree$C, reaching stressful conditions for the \conusplus~detector operation. For this reason, an air cooling (AC) system was installed to keep a stable temperature around 22$\degree$C and reducing variations below 2$\degree$C. Finally, the air-borne radon concentration in the room turned out to have an average of (110$\pm$7)~Bq~m$^{-3}$ and variations up to (80$\pm$5)~Bq~m$^{-3}$. 

During reactor operation relatively strong vibrations are produced close to the reactor core. These small movements of the detector and its surrounding components can induce microphonic events~\cite{Bonet:2020ntx}. For this reason, the vibrations were evaluated in all five positions of figure~\ref{conusplus_room} and were compared to the ones observed at the Max-Planck-Institut f\"ur Kernphysik (MPIK) under laboratory conditions. The vibrations were measured up to 3~kHz with a Kistler LabAmp Type 5165A, which is a low-noise charge amplifier for dynamic signals with an integrated data acquisition system. The device is based on three piezoelectric sensors (one per dimension) with a sensitivity of 10~mV/g and a sampling rate of 10~MHz.  No significant vibration increases were found in the vertical axis, while in the horizontal plane the amplitudes were six times larger than under laboratory conditions. However, their contribution has still a negligible impact on the overall \conusplus~detector performance. 

\section{Gamma-ray background and surface contamination}
\label{sec:2}

\subsection{Gamma-ray background}
\label{sec:2p1}

The $\gamma$-ray background in room ZA28R027 was studied with the low background p-type coaxial HPGe CONus RADiation (CONRAD) detector~\cite{Hakenmuller:2023yzb} at the five different positions that are indicated in figure~\ref{conusplus_room}. The spectrum below 2700~keV$_{ee}$ is expected to be dominated by natural radioactivity. Above this energy one expects almost exclusively $\gamma$-ray lines that are induced by neutron-captures in surrounding materials which can reach energies up to 10~MeV. The method allows for the identification of the isotope production in the surrounding environment and from the $\gamma$-line count rate the reactor neutron fluence can be indirectly estimated. 

\begin{figure}
    \centering
    \includegraphics[width=0.48\textwidth]{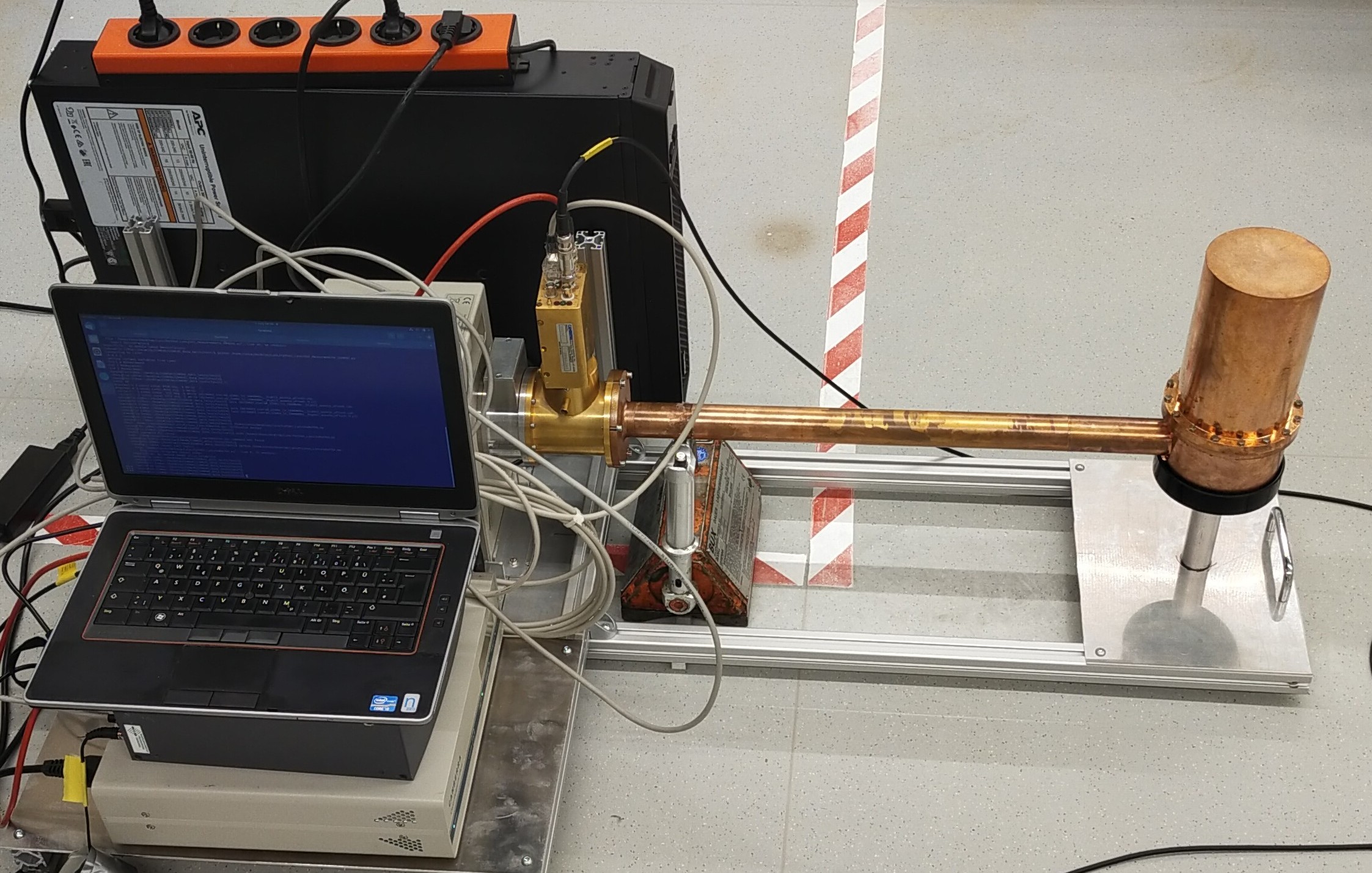}
    \caption{CONRAD setup including the HPGe detector, the UPS, the LYNX DAQ, the portable PC and the aluminum support structure.}
    \label{setup_CONRAD}
\end{figure}

CONRAD has an active mass of (1.90$\pm$0.15)~kg that allows to detect high energetic $\gamma$-lines up to 11~MeV$_{ee}$. The trigger efficiency was measured to be constant in the whole energy range using different pulser scans. Several energy calibrations were performed with a $^{60}$Co source before and after operation at reactor, ensuring the energy resolution and energy scale stability of the detector during the whole background determination campaign. As indicated in~\cite{Hakenmuller:2023yzb}, due to a small non-linearity within the energy scale the energy regions below 2.7~MeV$_{ee}$ and above 3.5~MeV$_{ee}$ are calibrated independently with two linear functions. The data collection was performed with a LYNX DAQ system, which also provides a stable high and low voltage supply to the Ge detector. Additionally, an uninterruptible power supply system (UPS) was installed to avoid power cuts during the operation at KKL. All these devices were installed on an aluminum support structure for detector transport and to reduce the risk of surface contamination (sc. Sec.~\ref{sec:2p2}). The setup used for the $\gamma$-ray background measurement at KKL is shown in figure~\ref{setup_CONRAD}.

\begin{figure*}[ht]
    \centering
    \includegraphics[width=0.95\textwidth]{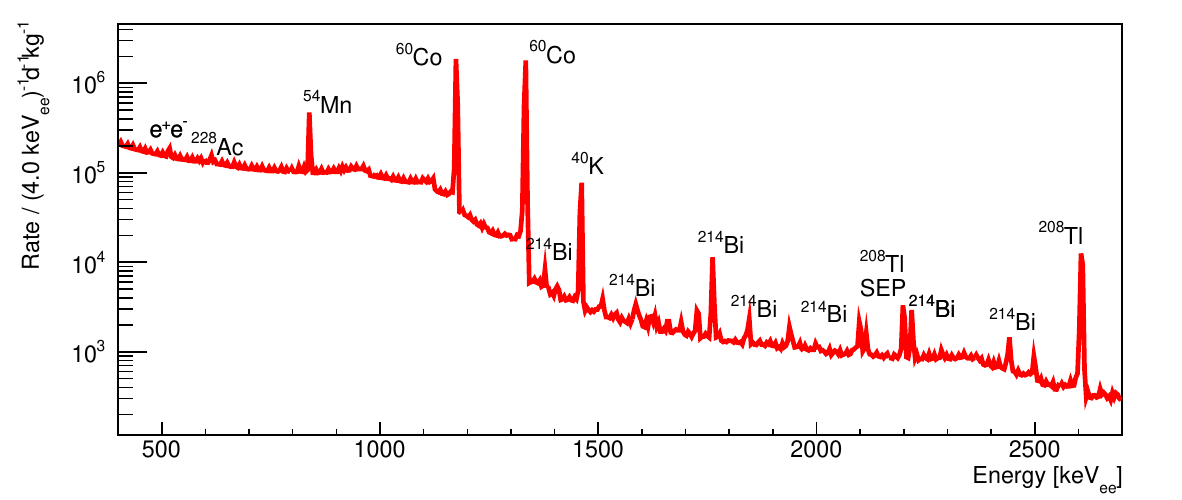}
    \caption{Gamma-ray energy spectrum below 2.7~MeV as measured with the non-shielded CONRAD detector inside ZA28R027. The spectrum is dominated by natural radioactivity emitted from the surrounding structural materials. The decaying isotopes are mentioned for the most intense $\gamma$-lines.}
    \label{conrad_spectra_low}
\end{figure*}

From the $\gamma$-ray measurement conducted with CONRAD inside ZA28R027 one can conclude that natural radioactivity indeed dominates the spectrum below 2.7~MeV$_{ee}$. The spectrum from a 1~day lasting measurement in position 2 is displayed in figure~\ref{conrad_spectra_low}. Regarding the other four positions, the contributions from the Th and U decay chains and $^{40}$K are present with similar $\gamma$-line intensities. However, the most intense $\gamma$-lines are from $^{60}$Co decays. Their intensities amount to $2\cdot10^5$~counts~d$^{-1}$kg$^{-1}$ in position 2, and vary up to factor of three in the other positions. 

\begin{figure*}
    \centering
     \includegraphics[width=0.95\textwidth]{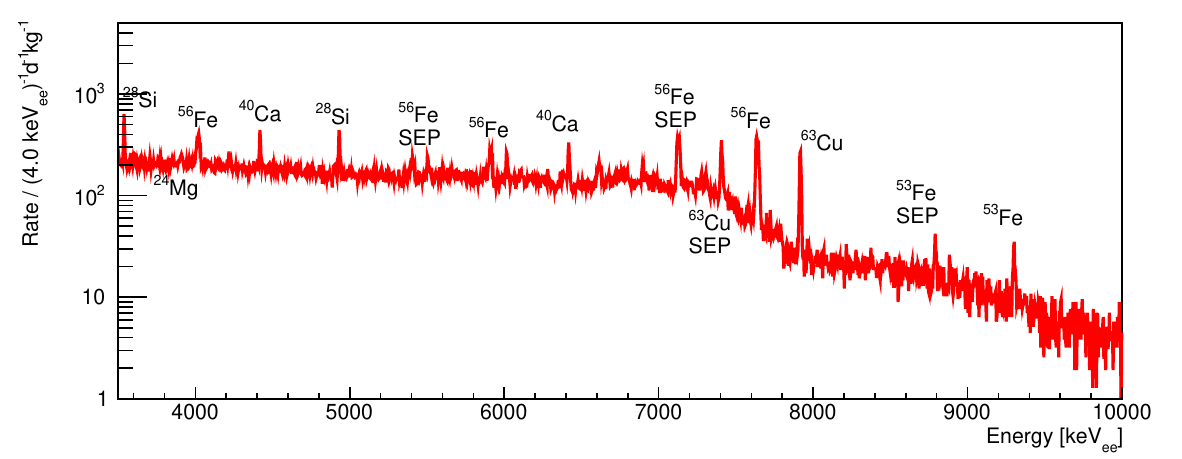}
    \caption{Gamma-ray energy spectrum above 3.5~MeV as measured with the non-shielded CONRAD detector inside ZA28R027. The spectrum is dominated by $\gamma$-ray lines, that are induced by neutron-capture radioisotopes and whose intensities are correlated with the reactor thermal power. The decaying isotopes are mentioned for the most intense $\gamma$-lines~\cite{IAEE:2007}.}
    \label{conrad_spectra_high}
\end{figure*}

The spectrum above 3.5~MeV$_{ee}$ obtained in a 1~day lasting measurement in position 2 in ZA28R027 is shown in figure~\ref{conrad_spectra_high}. All visible $\gamma$-lines are neutron-induced and correlated with the neutrons, that escape the reactor core, and thus with the reactor thermal power. The accompanying $\gamma$-radiation originates from neutron captures in the structural materials of the reactor. In particular, they originate from $^{28}$Si, the most abundant isotope in silicon (92.2\%) and from $^{40}$Ca (94.8\%), which are typically present in the Portland cement~\cite{Bediako:2015} and aggregates used in the KKL reinforced concrete. 
Additionally, several $\gamma$-lines from $^{53}$Fe and  $^{56}$Fe are visible, both attributed to neutron captures in the metallic parts of the reactor structure. Not only the full-energy peak (FEP) lines, but also single and double escape peaks (SEP and DEP) are observed, where one or both of the $\gamma$-rays from e$^{-}$e$^{+}$ pair production and subsequent annihilation within the HPGe diode can escape the detection volume. In general, the integral rate above 2.7~MeV$_{ee}$ is reduced about 26 times compared to ZA408 at KBR, decreasing from 155 to 6~counts~s$^{-1}$~kg$^{-1}$.

In order to understand this difference in $\gamma$-radiation at KKL and KBR, the count rates per day, kilogram and GW are summarized in table~\ref{reactor_gamma_values} for the main reactor correlated $\gamma$-lines observed above 3.5~MeV. Next to the main experimental rooms ZA28R027 at KKL and ZA408 at KBR~\cite{Hakenmuller:2019ecb}, the Ex-HPU-B position at KKL was investigated (see figure~\ref{power_plant} for more details). The $\gamma$-line energies and branching ratios were extracted from~\cite{IAEE:2007}. The count rates were determined by a counting method except for the double peak structure of $^{57}$Fe at 7631~keV and 7641~keV, which was determined following the method described in~\cite{Hakenmuller:2019ecb}. The systematic uncertainty on the CONRAD mass is the dominant factor for the count rate estimation, considering the energy calibration a second order effect. 

\begin{table*}
\caption{Main neutron-induced $\gamma$-lines over 3.5~MeV$_{ee}$ identified in CONRAD's energy spectra that were measured at KKL and KBR during reactor on periods. At KKL two measurement positions were considered: the \conusplus~room ZA28R027 and the Ex-HPU-B room. At KBR the measurement taken in the CONUS room ZA408 is reported. The production channels (PC), the energy and PC branching ratios (br) from literature are given as well~\cite{IAEE:2007}. The count rates are given in terms of [counts d$^{-1}$ kg$^{-1}$ GW$^{-1}$].   
Count rates marked with * were measured as one-peak structure and the single count rates were extracted as described in~\cite{Hakenmuller:2019ecb}.}
\centering
\setlength\extrarowheight{4pt}
\begin{tabular}{lccc}
\hline
PC / Energy [keV$_{ee}$] /BR[\%] & KKL: Ex-HPU-B & KKL: ZA28R027   & KBR: ZA408 \Tstrut\Bstrut\\
\hline
$^{53}$Fe(n,$\gamma$)$^{54}$Fe & (reactor structure) \Tstrut\Bstrut\\
\hline
8787 SEP & 573$\pm$45& 18.5$\pm$1.5  & 5.1$\pm$0.8 \\
9298 (100\%) & 707$\pm$56 & 19.7$\pm$1.6 &  5.9$\pm$0.9\Bstrut\\
\hline
$^{56}$Fe(n,$\gamma$)$^{57}$Fe & (reactor structure)\Tstrut\Bstrut\\
\hline
4217 (23.0\%) & 1896$\pm$149 & 78.5$\pm$6.2  & not visible \\
5920 (33.8\%)& 2504$\pm$190 & 95.1$\pm$7.5 &  not visible \\
6018 (34.8\%)& 2787$\pm$220  & 98.0$\pm$7.4 & not visible \\
7120 SEP & 6974$\pm$596*  & 285$\pm$23* & not visible \\
7135 SEP & double peak & double peak & double peak\\
7278 (20.7\%)& 1544$\pm$122  & 80.5$\pm$6.4&  6.1$\pm$0.9 \\
7631 (100\%)&  8717$\pm$735* & 363$\pm$29* &  72$\pm$11* \\
7646 (86.2\%)& double peak & double peak &  double peak\Bstrut\\
\hline
$^{63}$Cu(n,$\gamma$)$^{64}$Cu  & (HPGe cryostat) \Tstrut\Bstrut\\
\hline
7406 SEP & 1992$\pm$157 & 228$\pm$18&   14.3$\pm$2.1 \\
7638 (48.9\%)& 995$\pm$79* &120$\pm$9*  &   8.3$\pm$1.2*\\
7916 (100\%)& 2034$\pm$161 & 245$\pm$19  &   15.6$\pm$2.3\Bstrut\\
\hline
$^{28}$Si(n,$\gamma$)$^{29}$Si &(concrete \conusplus~room) \Tstrut\Bstrut\\
\hline
3539 (100\%)&   not visible & 276$\pm$22  &  not visible \\
4934 (93.3\%)&  not visible & 213$\pm$17  &   not visible \\
6379 (16.0\%)&  not visible& 19.9$\pm$1.6 &  not visible \\
7199 (10.0\%)& not visible & 6.8$\pm$0.5  &  not visible \Bstrut\\
\hline
$^{40}$Ca(n,$\gamma$)$^{41}$Ca &(concrete \conusplus~room) \Tstrut\Bstrut\\
\hline
4418 (17.1\%)&   not visible & 105$\pm$8  &  not visible \\
6419 (43.5\%)& not visible & 181$\pm$14  &  not visible \Bstrut\\
\hline
$^{16}$O(n,p$\gamma$)$^{16}$N  &(reactor cooling system) \Tstrut\Bstrut\\
\hline
5617 SEP & 7143$\pm$564 &not visible & 26301$\pm$3945 \\
6128 (67\%)& 12652$\pm$998&not visible  &  44782$\pm$6774\\
7115 (4.9\%)&  2526$\pm$199 &not visible  &  5314$\pm$797 \Bstrut\\
\hline
\label{reactor_gamma_values}
\end{tabular}
\end{table*} 

The $\gamma$-lines related to $^{28}$Si and $^{40}$Ca are only observed in the ZA28R027 room. In the case of Ex-HPU-B it can be explained with a lack of concrete around this location. In the case of ZA408 ~\cite{Hakenmuller:2019ecb} it can be explained with the usage of another type of reinforced concrete that is e.g. not silica-fume based. On the contrary, the $^{16}$N $\gamma$-radiation, which is produced in (n,p) reactions on $^{16}$O in the water of the cooling cycle, is relatively intense at ZA408, while at Ex-HPU-B it is partially suppressed and at ZA28R027 fully suppressed. The reason is, that the two KKL positions are more distant from their cooling cycle compared to the KBR position, whose distance of the CONUS detector to its primary cooling cycle was only 3.8~m~\cite{Hakenmuller:2019ecb}.
For the KKL locations, the highest contribution comes from the Fe isotopes, with values almost 13 and 250 times larger than at KBR in the ZA28R027 and Ex-HPU-B positions, respectively . This difference can be achieved with a greater amount of steel or a higher neutron flux. However, looking at the $^{63}$Cu isotope produced in the CONRAD copper cryostat, the second option can be confirmed as the main effect. A 16 times larger count rate was measured in ZA28R027 compared to KBR, which increased up to 140 times in the closer position. This suggests a larger neutron flux at KKL. This result is in agreement with the direct neutron fluence measurements reported in Sec.~\ref{sec:4}.

\subsection{Surface contamination}
\label{sec:2p2}

The {\it in-situ} measurements of the $\gamma$-ray background performed with the CONRAD detector (Sec. \ref{sec:2p1}) did not allow to distinguish the two contributing parts: nuclear radiation originating from the bulk of the surrounding materials and fluids (i.e. decays inside the concrete, steel structure or water cooling cycles) and radiation attributed to volatile surface contamination (i.e. dust particulates with radioisotopic contamination). The latter one can have stronger local and temporal variations and an accidental, even smaller cross-contamination of the intrinsically ultra low background CONUS shield and detector parts during their assembly could easily occur and have an unwanted enhancement of the background.\\

In order to quantify the level of surface contamination and its potential impact on CONUS and \conusplus~data, we performed a series of wipe tests of floor and wall surfaces at both experimental sites ZA28R027 in KKL and ZA408 in KBR before/after adopting cleaning procedures. The wipe tests were performed with a water-soap mixture over an area of 100\,cm$^2$. An extraction efficiency of 10\% is considered in the follow. Wipe tests of shield and detector components were further performed before/after assembly at KBR and 5 years later during disassembly of the experiment to understand the risk of accidental cross-contamination and the effectiveness of counteractions.\\
82 out of 105 performed wipe tests were evaluated by means of $\gamma$-ray spectroscopy using the low background Ge detectors Bruno and Corrado \cite{BUDJAS2009706} operated at the Low Level Laboratory of MPIK. In this case the measurements lasted 3.1 days on average, accumulating a total measurement livetime of 253 days. For a subset of 56 wipes, 30-minutes lasting $\alpha$/$\beta$ counting measurements using an LB790 proportional counter system by Berthold Instruments were conducted.\\

\begin{figure*}
    \centering
     \includegraphics[width=1.0\textwidth]{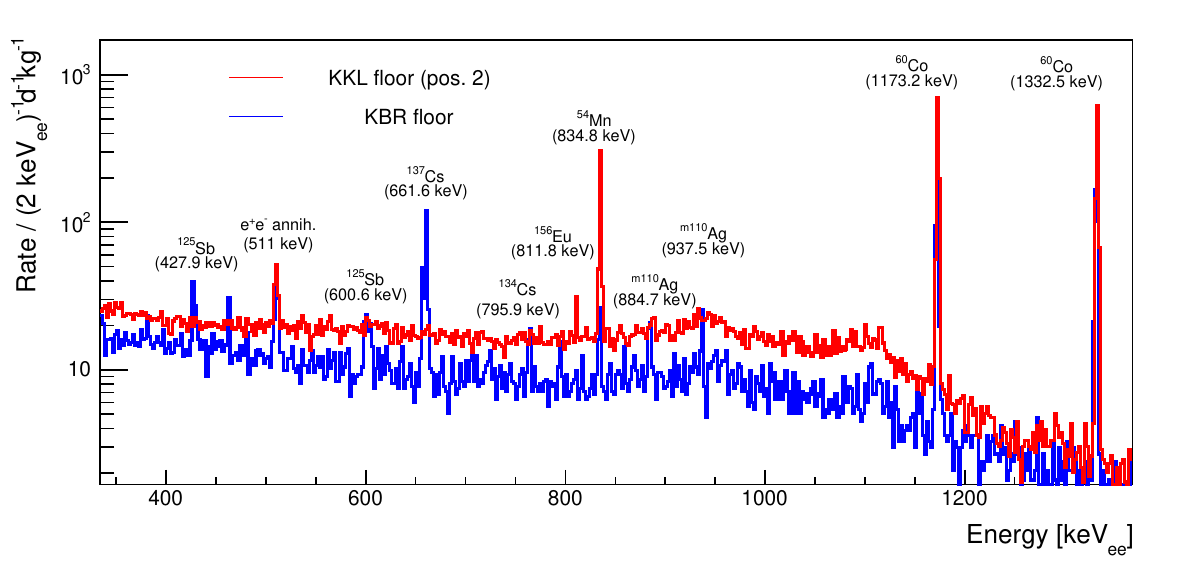}
    \caption{Energy spectra of two representative wipe tests taken at the CONUS and \conusplus~experimental sites, as measured with low background Ge $\gamma$-ray spectroscopy in the Low Level Laboratory of MPIK.}
    \label{floor_surfacecontam_kkl_kbr}
\end{figure*}

\begin{table*}[bht]
\begin{center}
\begin{footnotesize}
\caption{List of artificial radioisotopes found in surface contamination of the CONUS and \conusplus~experimental sites within the safety containment of the nuclear power plants Brokdorf (KBR) and Leibstadt (KKL) after $\sim$35 and $\sim$40 years of reactor operation. Their origin and most probable production mechanisms are from \cite{RadiologicalInventory98,DELMORE20111008}. As expected, the nuclide-vectors are reactor-dependent. 
\label{reactor_surfacecontamination_isotopes} }
\begin{tabular}{lllllll}
\hline
nuclide	&	decay mode	&	half-life	&	production channel	&	origin and presence	&	KKL	&	KBR  	\\
\hline
$^{60}$Co	&	$\beta^-$	&	5.27\,y	&	$^{59}$Co(n,$\gamma$)$^{60}$Co	&	$^{59}$Co: trace constituent in carbon and  &	x	&	x	\\
	&	&	 &	&stainless steel structural parts 	&	 		&	\\
$^{54}$Mn	&	$\beta^-$, $\beta^+$ &	312.3\,d	&	$^{54}$Fe(n,p)$^{54}$Mn 		&	$^{54}$Fe: present in construction materials, e.g.&	x	&	x	\\
	&		&	 	&	 			&	pressure vessel, fuel support, primary circuit&		&		\\
$^{137}$Cs	&	$\beta^-$	&	30.07\,y	&	--			&	$^{137}$Cs: progeny of the fission products $^{137}$Xe,  &-	&	x	\\
	&		&		&				&	$^{137}$I, $^{137}$Te leaking to primary coolant&	&	 	\\

$^{134}$Cs	&	$\beta^-$	&	2.06\,y	&	$^{133}$Cs(n,$\gamma$)$^{134}$Cs	&	$^{133}$Cs: progeny of the fission product $^{133}$Xe   &	-	&	x	\\
	&		&	 	&				&	leaking through fuel rods to primary coolant&		&	 	\\	
 $^{m110}$Ag	&	$\beta^-$	&	249.79\,d	&	$^{109}$Ag(n,$\gamma$)$^{m110}$Ag	&	$^{109}$Ag: present in Ag-In-Cd control rods and &	-	&	x	\\
	&		&	 	&				&	sealing the head of the reactor pressure vessel&		&	\\
 $^{156}$Eu	&	$\beta^-$	&	15.19\,d	&	$^{153}$Eu(3n,$\gamma$)$^{156}$Eu	&	$^{153}$Eu: Presence of rare earth elements in &	x	&	-	\\
	&		&	 	&				&	e.g. reactor graphite and bioshield concretes&		&		\\
 $^{125}$Sb	&	$\beta^-$	&	2.76\,y	&	$^{124}$Sb(n,$\gamma$)$^{125}$Sb	&	$^{124}$Sb: present in irradiated cladding &	-	&	x	\\
	&		&	 	&				&	 or in moderator tubes fabricated of Zr alloys	 &		&		\\
\hline
 \end{tabular}
\end{footnotesize}
\end{center}
\end{table*}

\begin{table*}[bht]
\begin{center}
\begin{footnotesize}
\caption{Ge-spectroscopic measurements of $\gamma$-line intensities from wipe tests performed in the experimental sites of CONUS at KBR and of \conusplus~at KKL. The count rates are reported for $\gamma$-lines in italic letters and are expressed in units of [counts d$^{-1}$ kg$^{-1}$]. The KKL floor positions coincide with the ones in Figure \ref{conusplus_room}, while the wall positions are from the walls closest to the \conusplus~shield setup. Entries denoted with `n.v.' are negligible / below the detection threshold of the used low background Ge detector. 
\label{reactor_surfacecontamination_results} }
\begin{tabular}{llllllllll}
\hline
\multicolumn{1}{l}{nuclide} &
\multicolumn{1}{l}{$\gamma$-lines / keV} &
\multicolumn{1}{l}{KBR floor} &
\multicolumn{4}{c}{KKL floor} &
\multicolumn{3}{c}{KKL	wall} \\ 
           &		&		&	Pos. 1	&	Pos. 2	&	Pos. 3	&	Pos. 4	&	Pos. 1	&	Pos. 2	&	Pos. 3	\\
\hline
$^{60}$Co	&	{\it 1332.5}, 1173.2	&	286	&	1328	&	845	&	597	&	166	&	710.1	& 267.6	&	199.4	\\
$^{54}$Mn	&	{\it 834.8}	&	25	&	35	&	326	&	56	&	10	&	6.3	&	0	&	28.3	\\
$^{137}$Cs	&	{\it 661.7}	&	141	&	n.v.	&	n.v.	&	n.v.	&	n.v.	&	n.v.	&	n.v.	&	n.v.	\\
$^{134}$Cs	&	{\it 795.9}, 604.7	&	14	&	n.v.	&	n.v.	&	n.v.	&	n.v.	&	n.v.	&	n.v.	&	n.v.	\\
$^{m110}$Ag	&	{\it 884.7}, 763.9, 937.5, 1384.3&	32	&	n.v.	&	n.v.	&	n.v.	&	n.v.	&	n.v.	&	n.v.	&	n.v.	\\
$^{156}$Eu	&	{\it 811.8}, 1076.0, 1079.2, 2269.9	&	n.v.	&	9.5	&	19.1	&	3.5	&	7.7	&	n.v.	&	n.v.	&	n.v.	\\
$^{125}$Sb	&	{\it 427.9}, 380.5, 600.6	&	37	&	n.v.	&	n.v.	&	n.v.	&	n.v.	&	n.v.	&	n.v.	&	n.v.\\
\hline
 \end{tabular}
\end{footnotesize}
\end{center}
\end{table*}

As summarized in Table \ref{reactor_surfacecontamination_isotopes}, the high resolution Ge energy spectra of the wipes revealed the presence of different surface-soluble isotopes, whose characteristic composition depends on the reactor sites. Wipe tests from KKL contain mainly $^{60}$Co and $^{54}$Mn, but also traces of $^{156}$Eu. On the contrary, wipes from KBR contain mainly $^{60}$Co and $^{137}$Cs, with additional traces of $^{54}$Mn, $^{134}$Cs, $^{m110}$Ag and $^{125}$Sb. All of them are either progenies of fission products in the fuel rods or are produced in reactor materials via neutron activation. They are released due to corrosion/erosion processes and microscopic leaks continuously or during outages and decommissioning phases. As confirmed by our $\alpha$/$\beta$ counter measurements, among the detected isotopes there are only $\beta$- and no $\alpha$-emitters.\\
Figure \ref{floor_surfacecontam_kkl_kbr} depicts energy spectra of two representative wipe tests of the floor surface taken in room ZA28R027 at KKL (pos. 2 in Figure \ref{conusplus_room}) as well as in room ZA408 at KBR, as measured by means of Ge $\gamma$-spectroscopy. The  $\gamma$-line intensities found in several wipe tests of floor and wall surfaces at the two places are reported in Table \ref{reactor_surfacecontamination_results}. On one side all intensities are relatively small, on the other side the $^{60}$Co isotope dominates especially within the KKL wipe tests; the overall determined mean count rate of the related 1332.5\,keV $\gamma$-line is (425$\pm$160) and (275$\pm$215)\,counts d$^{-1}$ for KKL floor and walls, respectively. From the $\beta$-counting method, which sums over all isotopic contributions, one obtains an averaged surface activity of (48$\pm$16) versus (36$\pm$15)\,mBq~cm$^{-2}$. In a few cases, hot spots were localized on floor and wall surfaces with activities $\sim$10 times above average. Walls are in general cleaner, and the $^{156}$Eu isotope is almost exclusively found on floor samples.\\

From wipe tests of equipment used in this work, we observed that it can be cross-contaminated, specifically, when setups such as electrical cryocoolers and DAQs, that include ventilation systems, are placed close to the floor over a period of several weeks. In order to quantify the impact of a cross-contamination on the background budget of \conusplus~detectors, we placed the wipe of the KKL floor pos. 2 on the top outer surface of Corrado's end cap. This low background detector has an active mass of 845\,g, which is similar to the one of the CONUS detectors (sc. Table 1 in \cite{Bonet:2020ntx}). For the same $\gamma$-emitter this leads to similar Peak-to-Compton ratios and thus comparable spectral shape contributions over a wide energy range. It was found, that the background rate of Corrado in the region (100,440)\,keV increases by 2400\,counts d$^{-1}$kg$^{-1}$. Adding this contribution to the background rates measured at KBR with the C1, C2 and C3 detectors (sc. Table 8 in \cite{Hakenmuller:2023yzb}), this would translate into a background increase of factor 4.5 to 8.7. In other regions of interest at the keV scale, a similar background increase is expected. This demonstrates the difficulty to carry out an ultra low background rare-physics-process experiment at such locations demanding proper measures. In CONUS we applied the following counteractions. First, floors and walls were cleaned with water-soap mixture allowing to reduce the surface contamination by up to a factor $\sim$5. Second, during transport of the setup all parts were wrapped into plastic foil. Third, assembly took place under regular exchange of gloves and partly within a protection tent with slight overpressure. Wipe tests of the dissembled CONUS setup at KBR in December 2022 revealed that on this way the radiopurity of most shield and detector parts can be guaranteed over a period of at least 5 years. Only the lower part of the outer stainless steel cage showed some contamination. We found, that a proper cleaning with isopropanol can reduce such a surface contamination on smooth metallic surfaces very efficiently, while for rough plastic surfaces it is less efficient.

\section{Cosmic muon background}
\label{sec:3}

The muon flux in the room  ZA28R027 is evaluated with a small liquid scintillator detector designed and assembled at MPIK, which allows to measure muons up to 16~MeV. It consists of a cylindrical PTFE cell with a diameter of 5~cm and a height of 6~cm, resulting in an active volume of about 120~cm$^{3}$. An UV-transparent glass plate couples the cell to a 2" ET Enterprise 9954B photo-multiplier tube (PMT) which detects the scintillation light. Both are embedded inside a light tight aluminum housing. The cell is filled with Ultima Gold F from Perkin Elmer, a liquid scintillator chosen for its safer high flash point (150$^{\circ}$C), fast response, high light yield and good pulse shape particle discrimination (PSD) properties. The signals are acquired with a 250~MHz CAEN V1725 analog-to-digital converter (ADC) module controlled with the CAEN CoMPASS software. 

Before the measurements at KKL the detector was calibrated at MPIK with a $^{137}$Cs source ($\gamma$-line at 661~keV), and with the $\gamma$-lines at 1460~keV and 2615~keV from $^{40}$K and $^{208}$Tl, respectively, induced onsite by natural radioactivity. Because of the small volume of the cell, the Compton edge is used for energy calibration following the method described in~\cite{Bonhomme:2022xcg}. The presence of impurities, as oxygen, can degrade the light yield of liquid scintillators~\cite{Hua-Lin:2009gyx}. For this reason, several calibrations were performed during the background characterization campaign to ensure a stable light yield.   

\begin{figure}
    \centering
    \includegraphics[width=0.48\textwidth]{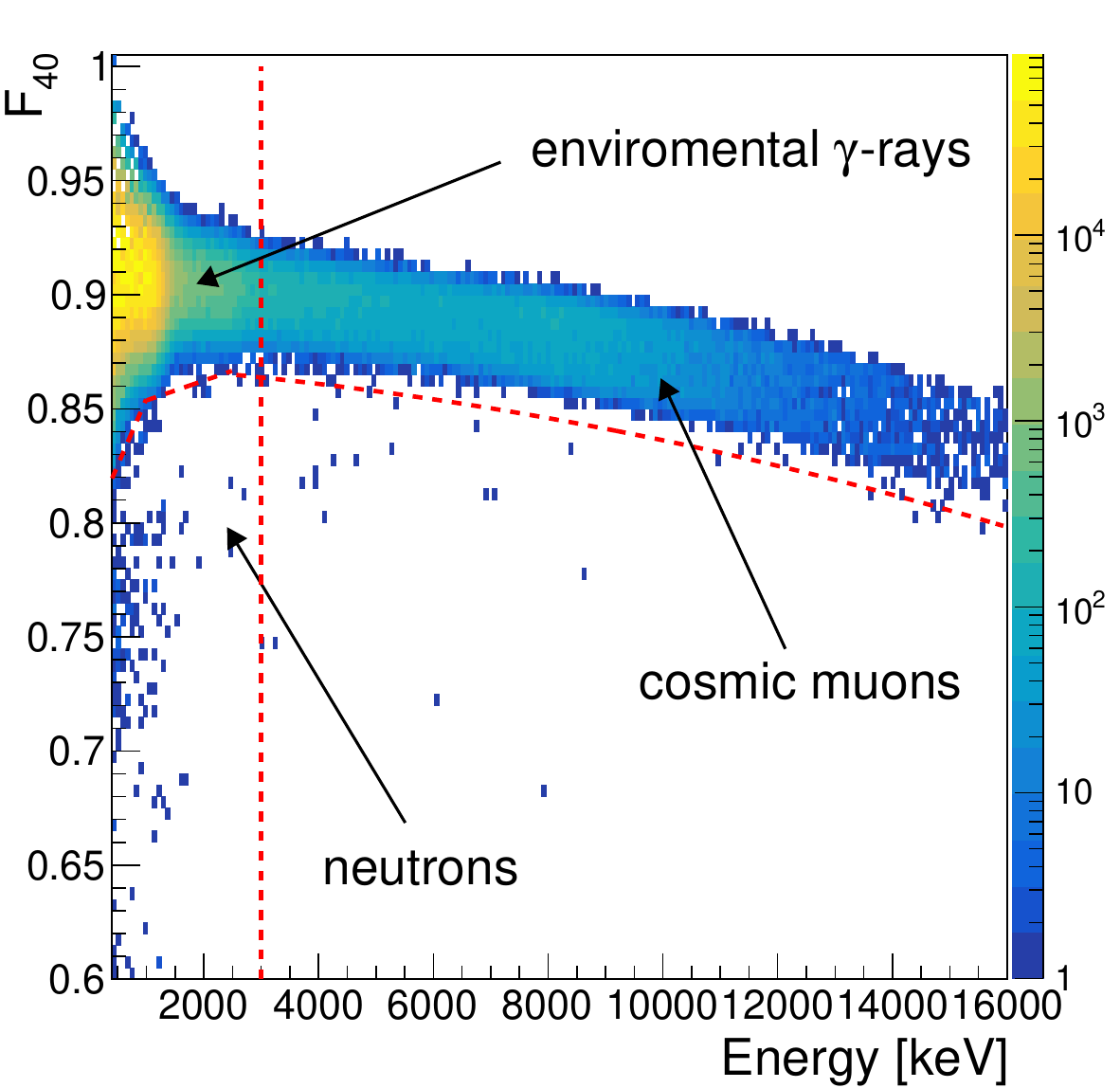}
    \caption{F$_{40}$ variable versus deposited energy as measured at KKL in reactor on time during one day. There are three distributions of events associated with neutrons, environmental $\gamma$-rays, and cosmic muons. A PSD cut and a 3~MeV energy threshold cut (two red dashed lines) are applied to select muon events.}
    \label{muon_psd2}
\end{figure}

The pulse shape analysis of the scintillation light can be used to identify and discriminate different incident particle types. The different ratios of charge integrals of a short gate window covering the beginning of the pulse (Q$_{fast}$) and a long gate period over the full signal (Q$_{tot}$), provides a powerful particle discrimination technique in liquid scintillators. A new variable, $F_{40}$, is defined as the fraction of the light detected in the first 40~ns of the pulse respect to the the full 1~$\mu$s signal length. The measured total integrated charge is represented with respect to the $F_{40}$ value in figure~\ref{muon_psd2} at KKL during a reactor off time period. There are two bands of events, separated by different $F_{40}$ values. The lower band with $F_{40}$ values in the range [0.5, 0.85] is associated mainly with neutrons, while the upper band in [0.85, 0.95] is due to electron recoil events. 

Two distributions of events can be differentiated in the electron recoil band. The first one at low energy below 2.7~MeV is consistent with $\gamma$-rays from environmental radioactivity. The second distribution consists of a narrow band of events from 3 to 16~MeV, consistent with muons produced in cosmic-ray events. Saturation effects decrease the $F_{40}$ values at high energy for muons. For this reason, a data-based PSD cut is developed using an $^{241}$AmBe neutron source,to remove high energy neutrons while accounting for saturation effects. The muon region (indicated by red dashed lines), is estimated applying this PSD cut and a 3~MeV energy threshold. The environmental $\gamma$-rays and the very small neutron contribution are efficiently suppressed with these selection cuts.

\begin{figure}
    \centering
    \includegraphics[width=0.48\textwidth]{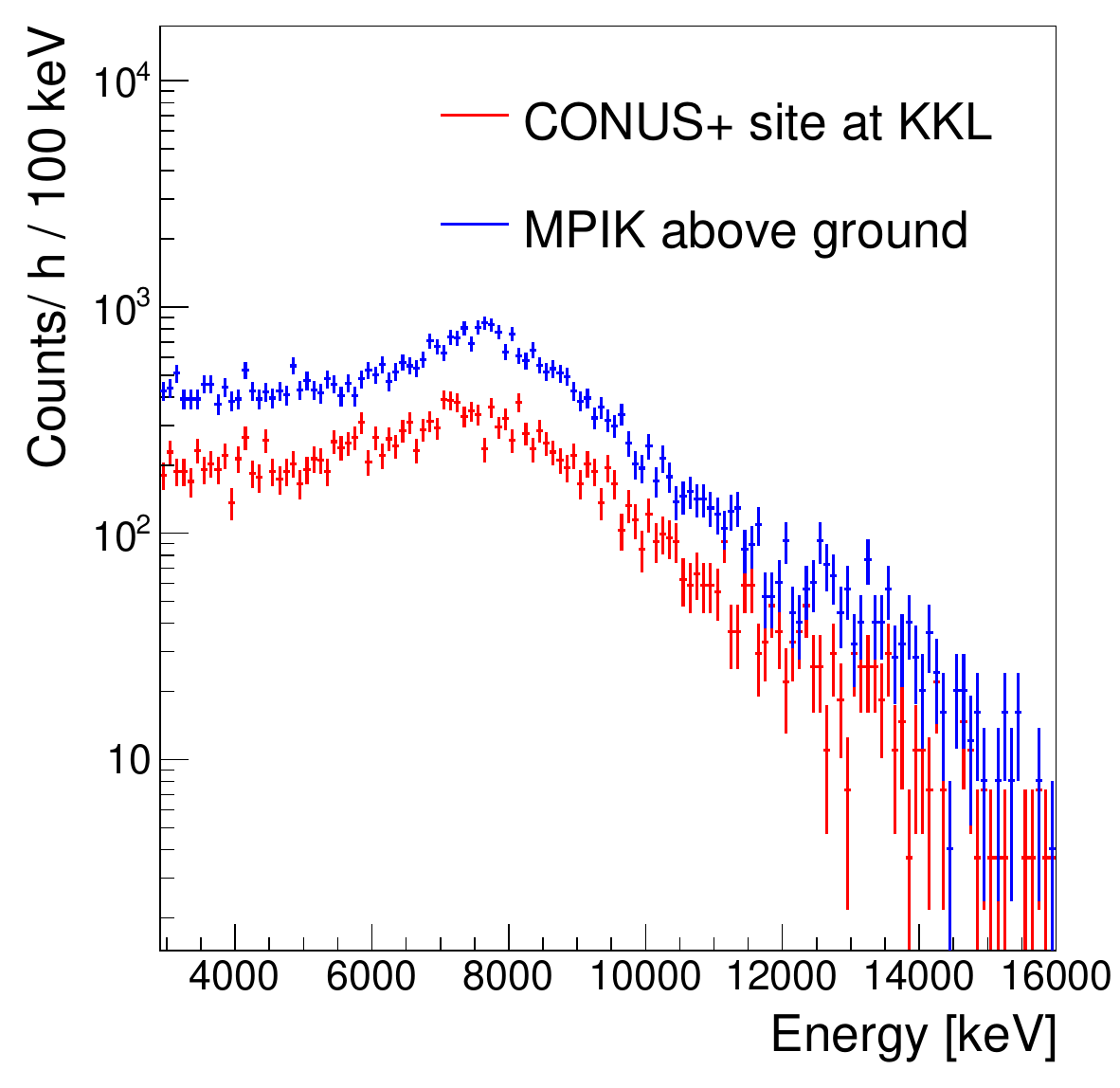}
    \caption{Muon energy spectra after selection cuts on surface (blue) and inside the \conusplus~room at KKL during reactor off time (red) normalize to one day exposure. Both spectra have the same shape with a 1.9 scaling factor, equivalent to an overburden of 7.4 m w.e.}
    \label{muon_spectra}
\end{figure}

The measured muon energy spectra at Earth's surface at MPIK campus (blue) and in the final \conusplus~position at KKL (red) are compared in figure~\ref{muon_spectra} after applying the selection cuts. The MPIK campus is situated at $\sim$350~m above sea level (m a.s.l.), which is very similar to room ZA28R027 at KKL located at 366~m a.s.l. Thus, similar muon fluxes are expected in both locations. The measurement at KKL was performed during reactor off time to avoid the impact of high energy $\gamma$-rays. The spectral shape is identical in both measurements with a bump at $\sim$~8~MeV produced by muons with a vertical incidence. 

A simple Geant4 simulation allows to deduce the muon flux from the scintillation detector rate. The muon flux measured at MPIK is (200$\pm$5)~muons~s$^{-1}$m$^{-2}$, in good agreement with the value reported in~\cite{HAUSSER1993223}. 
The muon flux is reduced to (107$\pm$3)~muons~s$^{-1}$m$^{-2}$ at KKL, which is 1.9 times smaller than at surface. This corresponds to an average overburden of (7.4$\pm$0.1)~m w.e. for the \conusplus~detector in ZA28R027 according to~\cite{HAUSSER1993223}. This value is in good agreement with the one estimated from the reactor's technical drawings in Sec.\,\ref{sec:2}. A muon flux reduction of $\sim$3\% is observed when the reactor drywell head is placed over the room, consistent with an overburden increased between 0.2-0.3~m~w.e. The cosmic-ray hadronic component is suppressed more than 2 orders of magnitude by the 7.4~m w.e  overburden. Additionally, the \conusplus~shield material over the Ge detectors provides an additional $\sim$~3~m w.e. of overburden, avoiding the cosmogenic activation of the \conusplus~HPGe detectors~\cite{HAUSSER1993223}. More details about the cosmogenic neutron flux are discussed in Sec.\,\ref{sec:4.2}.

The CONUS experimental location ZA408 at KBR had an average overburden of 24~m w.e.~\cite{Hakenmuller:2019ecb}. At this depth, the muon flux is reduced by 4.5~times with respect to Earth's surface. The number of muon-induced neutrons in Pb can be estimated for 7.4~m w.e. using~\cite{HAUSSER1993223} as $8.7\times 10^{-4}$ neutrons~s$^{-1}$m$^{-2}$, which is 2.3 times larger than in room ZA408 of KBR. For this reason, the \conusplus~shield was modified with the inclusion of a second inner muon veto system to improve the rejection efficiency of muon-induced background produced in the \conusplus~shield~\cite{CONUS:2024lnu}.



\section{Neutron fluence measurement at KKL}
\label{sec:4}

The neutron energy spectrum in room ZA28R027 at KKL was directly measured using the Extended Range Bonner Sphere Spectrometer (ERBSS) of the Paul Scherrer Institut (PSI) which consists of a set of thermal neutron sensors that are either operated barely or are surrounded by moderator spheres of different diameters and materials \cite{NEMUS2002}.
The mostly hydrogen-rich spheres moderate fast neutrons via multiple collisions and scattering processes and thereby turning them into detectable thermal neutrons. The thermal neutron sensors used in this work are spherical Centronic SP9 proportional counters filled with 2.3~bar of $^{3}$He and 1.2~bar of Kr. These counters detect thermalized neutrons via the reaction:\

\begin{equation}
\begin{aligned}
    \textnormal{n} + ^{3}\negmedspace\textnormal{He} \rightarrow ^{3}\negmedspace\textnormal{H} + \textnormal{p} + \textnormal{Q}, \ \ \textnormal{with Q=764~keV}.   
\end{aligned}
\label{he3_reaction}
\end{equation}

\begin{figure}
    \centering
    \includegraphics[width=0.48\textwidth]{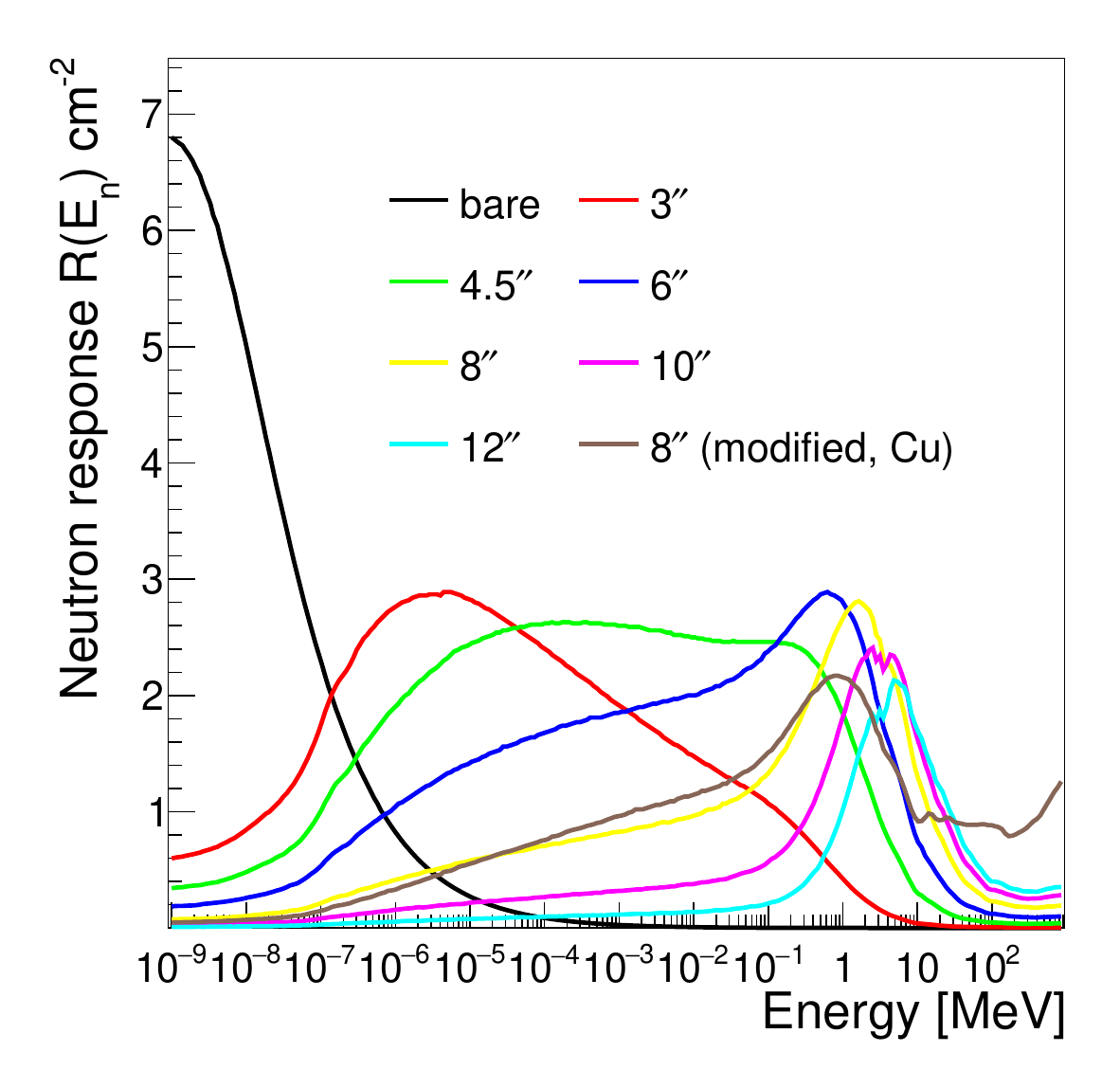}
    \caption{Neutron response functions of the Bonner spheres used for the measurement providing a wide energy sensitivity range from 10$^{-9}$~MeV to 10$^{3}$~MeV.}
    \label{sensitivity_spheres}
\end{figure}

While the bare sensors are sensitive to thermal neutrons, sensors enclosed in the differently configured PE spheres have a different energy-dependent response to neutrons. The spectrometer has been characterized via Monte Carlo methods \cite{GALEEV2021}. The results are verified through measurements in quasi-monoenergetic reference fields \cite{PIAF2009}.
The energy response of all these combinations as used for the measurements at KKL are shown in figure~\ref{sensitivity_spheres}. The peak of the neutron response function shifts to higher neutron energies as the size of the moderator sphere diameter increases. The response function dramatically increases for neutron energies above E$_{n}$$\sim$~50~MeV with the inclusion of a Cu shell. The latter case allows to check for the presence of high-energy cosmic-ray induced neutrons (E$_{n}$$\sim$~100 MeV~\cite{WIEGEL200252}).

Different methods can be used to deduce the spectral neutron distribution from measurements with ERBSS~\cite{REGINATTO2010}. However, the  deconvolution of a spectral neutron  distribution covering several orders of magnitude in energy from a few measurement points with linear dependent response functions can lead to different solutions which are in agreement with the data~\cite{REGINATTO2009}. A method suitable to analyze the presented measurements is Bayesian Parameter Estimation using a parameterized model~\cite{REGINATTO2006}. The analysis was carried out using Monte Carlo methods described in~\cite{JAGS2003}. The employed parametrization was optimized for spectral neutron distributions expected behind shielding of high-energy accelerators with a significant neutron component in the thermal and evaporation regions~\cite{TRS318}. \\

Prior to the neutron fluence detmination with the ERBSS system, the thermal neutron rate inside room ZA28R027 was evaluated with a bare SP9 counter in the first four positions indicated in figure~\ref{conusplus_room} during reactor on periods. A non-uniform neutron rate within the room was measured during reactor on, with values between 0.59 and 0.73~neutrons s$^{-1}$, following a 1/r$^{2}$ behaviour with r being the distance to the reactor core center. The neutron rate was also measured in the Ex-HPU-B position with values ranging from 4.9 to 6.4~neutrons s$^{-1}$, also following an 1/r$^{2}$ flux dependence. Due to the non-uniform neutron flux in the room during reactor on periods, the neutron fluence was evaluated as close as possible to the final \conusplus~location, exchanging the different spheres always in the same position.

\begin{figure}
    \centering
    \includegraphics[width=0.48\textwidth]{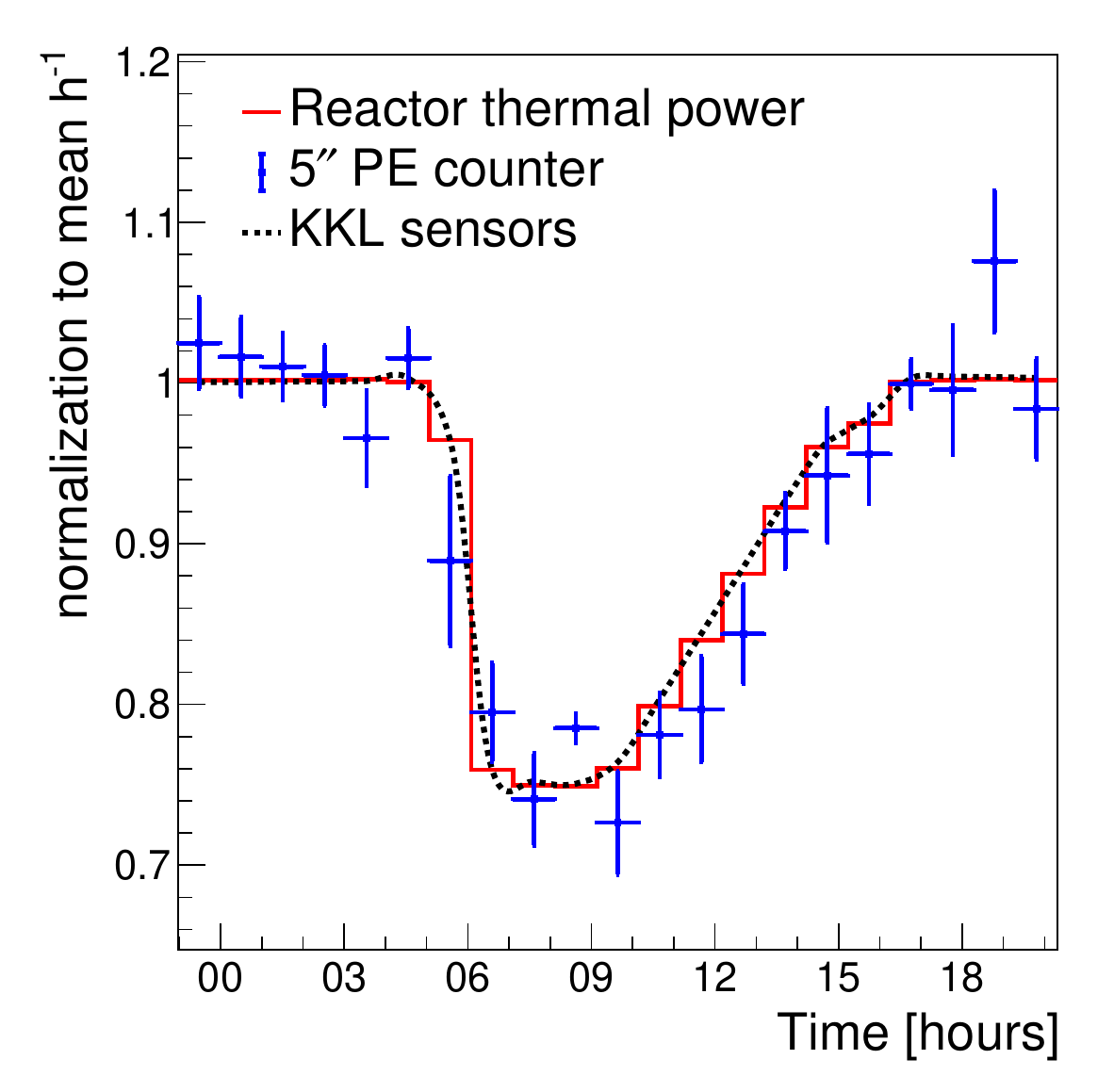}
    \caption{Evolution of neutron rate in 5\inches~sphere (blue), thermal power (red) and KKL neutron sensor (black) during reactor maintenance operations on December 3rd, 2023. All curves follow the same trend probing that most of the neutrons detected in the ZA28R027 room are produced by the reactor.}
    \label{neutron_rate_evolition}
\end{figure}

In addition, the neutron rate dependence with the thermal power was studied with a 5\inches\,sphere, which is mainly sensitive to a wide neutron energy range [10$^{-7}$, 10$^{-1}$]\,MeV (sc. figure~\ref{sensitivity_spheres}). In figure~\ref{neutron_rate_evolition} the evolution of the normalized count rate in the SP9 (blue marker) and the reactor thermal power (red solid) is shown, during a short reactor maintenance operation, in which the thermal power was reduced by approximately 25\%. The evolution of the normalized average count rate in the KKL neutron monitoring sensors (black dashed) is also included as a second reference. These consist of four NA-250 argon filled uranium fission chambers (22.5\%~$^{235}$U, 77.5\%~$^{234}$U). They are housed in steel dry-tubes and located in the moderator rich bypass region between the fuel elements. The neutron signals measured in room ZA28R027 follow the same trend observed in the other two curves also in terms of amplitude, probing that most of the detected neutrons are neutrons escaping the reactor core and which must be correlated with the thermal power.  For this reason, the neutron rate was monitored during the whole measurement campaign to ensure a stable neutron flux from reactor core.

The neutron fluence measurements at KKL were carried out using a subset of spheres to minimize the measurement time while still covering the complete energy range of the expected neutron field. The employed moderator configuration consisted of 6 polyethylene (PE) spheres of 3\inches, 4.5\inches, 6\inches, 8\inches, 10\inches~and 12\inches~in diameter, and one modified 8\inches~PE sphere containing a Cu shell of 1\inches~thickness. Five different SP9 proportional counters were used for the measurements at KKL with a Dextray ACHP96 pre-amplifier and a CAEN A1422 pre-amplifier. The signals were processed and acquired with the CAEN V1725 ADC module and the CAEN CoMPASS software, producing a pulse height spectra (PHS) for each measurement.  

\begin{figure}
    \centering
    \includegraphics[width=0.48\textwidth]{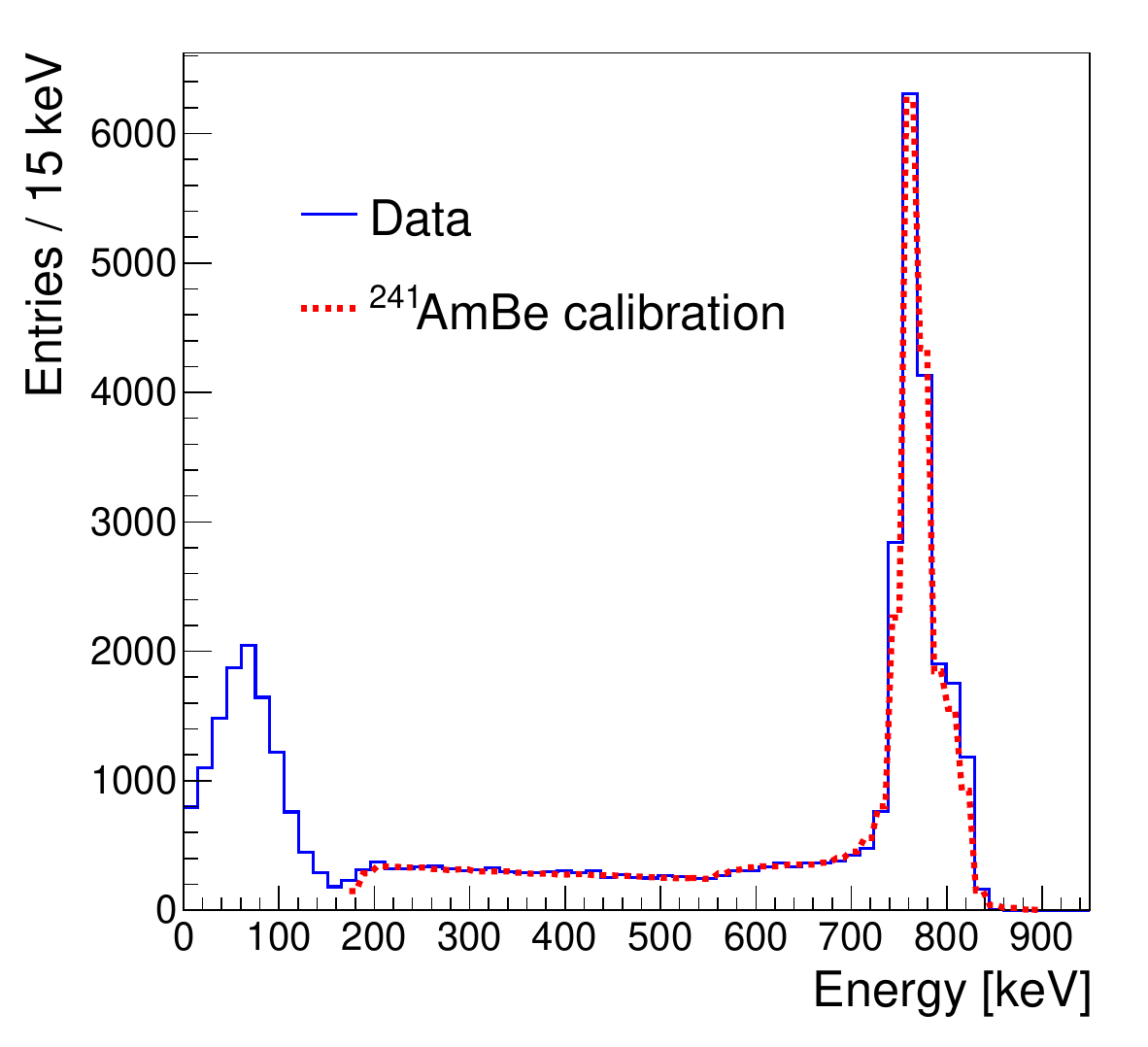}
    \caption{Pulse height spectrum of the bare $^{3}$He-filled SP9 proportional counter fitted to a reference spectrum (red) for the count rate determination. The reference spectrum was acquired with an $^{241}$AmBe source in the PSI verification stand.}
    \label{fit_neutron}
\end{figure}

Figure~\ref{fit_neutron} shows a typical PHS  measured at KKL (blue). Due to the low signal rate above threshold (down to 50~counts h$^{-1}$), the PHS was fitted with a measured reference PHS (red) to extract the neutron count rate. For this purpose, each SP9 detector was calibrated in the PSI verification stand using an $^{241}$AmBe source~\cite{GALEEV2021}. This also allows to correct for pressure difference between the volumetric equal counters, which could produce differences in the number of neutrons detected. The peak on the right side of the PHS corresponds to the full energy deposition of the reaction products in the gas (Q=764~keV), with a small distortion at higher energies produced by the pre-amplifier electronics. The continuous left part of the spectrum is formed by those particles which deposit only part of their energy in the gas and are absorbed in the wall of the detector. This is known as the `wall effect'~\cite{THOMAS200212}. The peak in the outermost left part is produced by environmental $\gamma$-rays, which are rejected with an energy threshold of $\sim$~160~keV. 

\subsection{Neutrons during reactor on time}
\label{sec:4.1}

\begin{figure}
    \centering
    \includegraphics[width=0.48\textwidth]{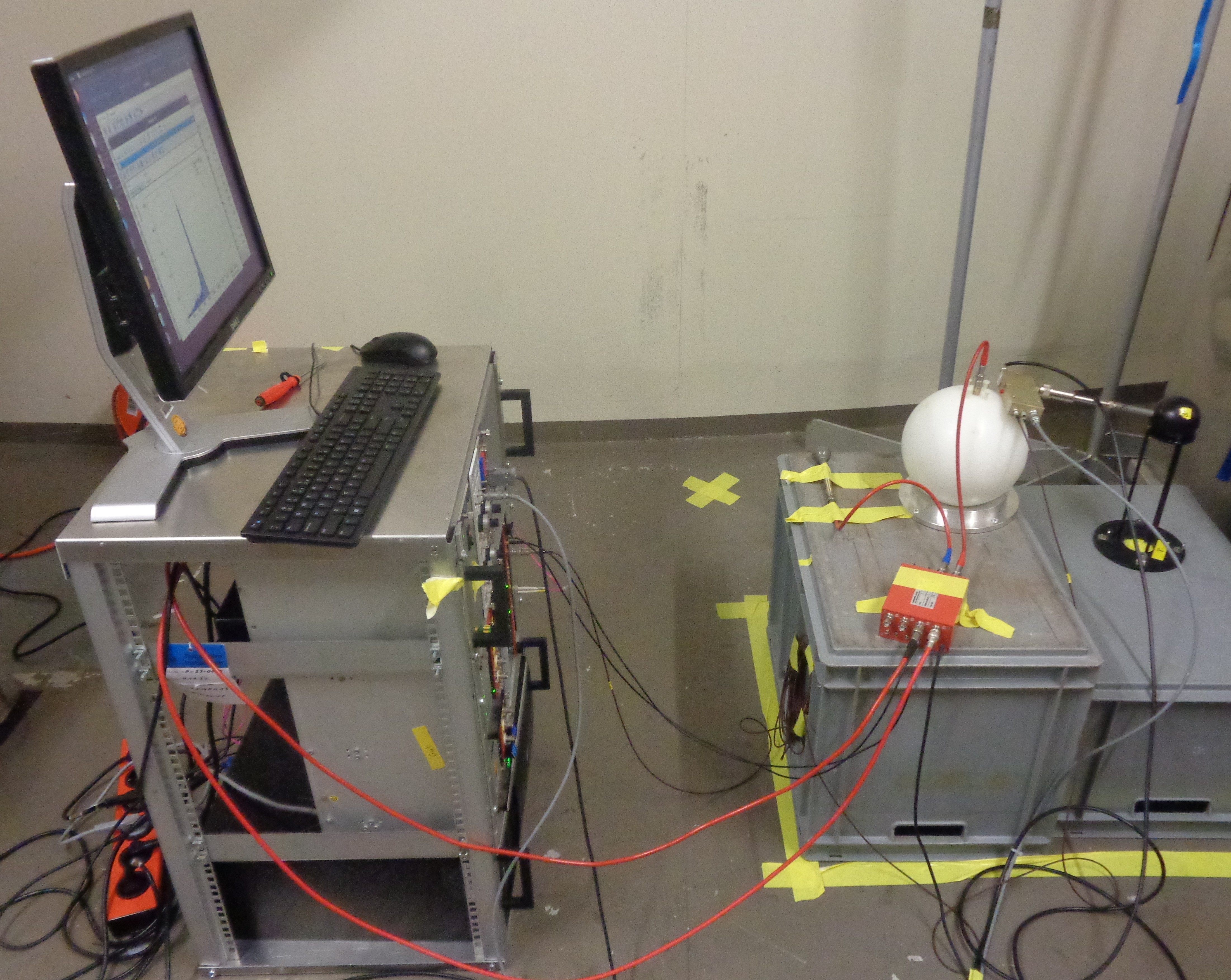}
    \caption{Experimental arrangement used during reactor on measurements. The bare detector and the 8\inches~PE on the left were used as a monitor of the neutron fluence rate. The moderator spheres were placed in position 5 of the ZA28R027 room, close to the final \conusplus~location (black sphere, right). Each moderator sphere was placed on this position one after another.}
    \label{spheres_ON}
\end{figure}

The data set during reactor on was acquired with the arrangement shown in figure~\ref{spheres_ON}, where the bare counter and all the different moderator spheres mentioned before, using the same SP9 counter, were placed one after another in position 5 in figure~\ref{conusplus_room}, which is close to the final \conusplus~shield-detector setup. The thermal neutron counters within the moderator spheres were at a height of 50~cm above ground. The measurement time for the different configurations amounted to 1-4 days due to the low count rate. The neutron flux was monitored during the whole campaign by a bare counter and an 8\inches~PE sphere placed at 30~cm from the other moderator spheres. The count rate measured in the latter two detectors was stable at a 3\% level, confirming the constant reactor thermal power during the whole measurement campaign. 

\begin{figure}
    \centering
    \includegraphics[width=0.48\textwidth]{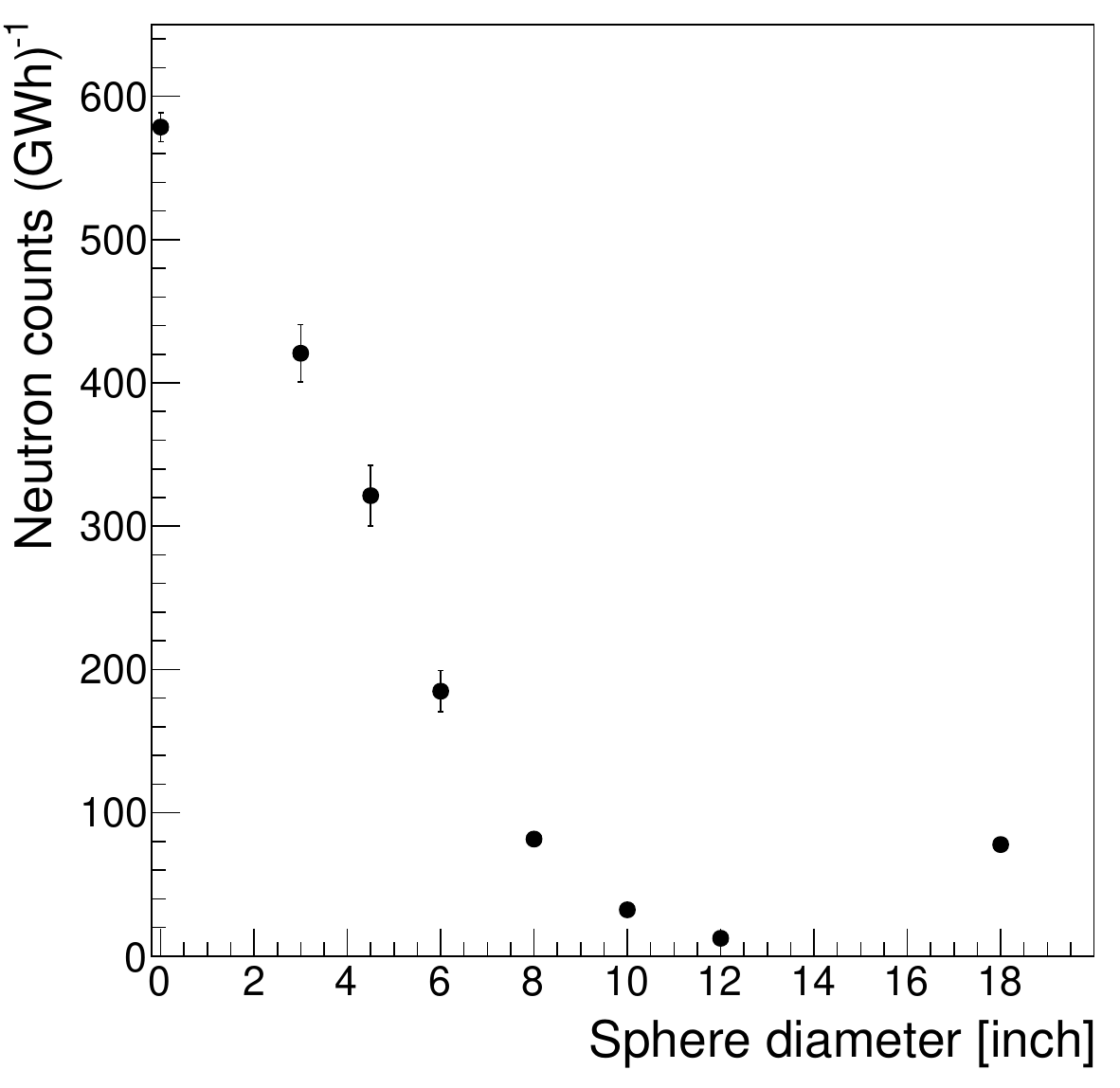}
    \caption{Neutron counts in the individual Bonner spheres during reactor on normalized to 1 GW h, as function of the sphere diameter. The data point at 0 corresponds to the bare detector. The data point at 18 corresponds to the 8\inches~modified sphere. The error bars are in some cases smaller than the size of the symbols.}
    \label{count_rate_ON}
\end{figure}

The neutron count rate for each sphere configuration is shown in figure~\ref{count_rate_ON} after normalizing by the reactor thermal power. The count rates were calculated fitting the PHS as described previously. The uncertainty of the fit is considered for each measurement individually. The bare counter was removed from the analysis due inconsistency with the other sphere configurations.

\begin{figure}
    \centering
    \includegraphics[width=0.48\textwidth]{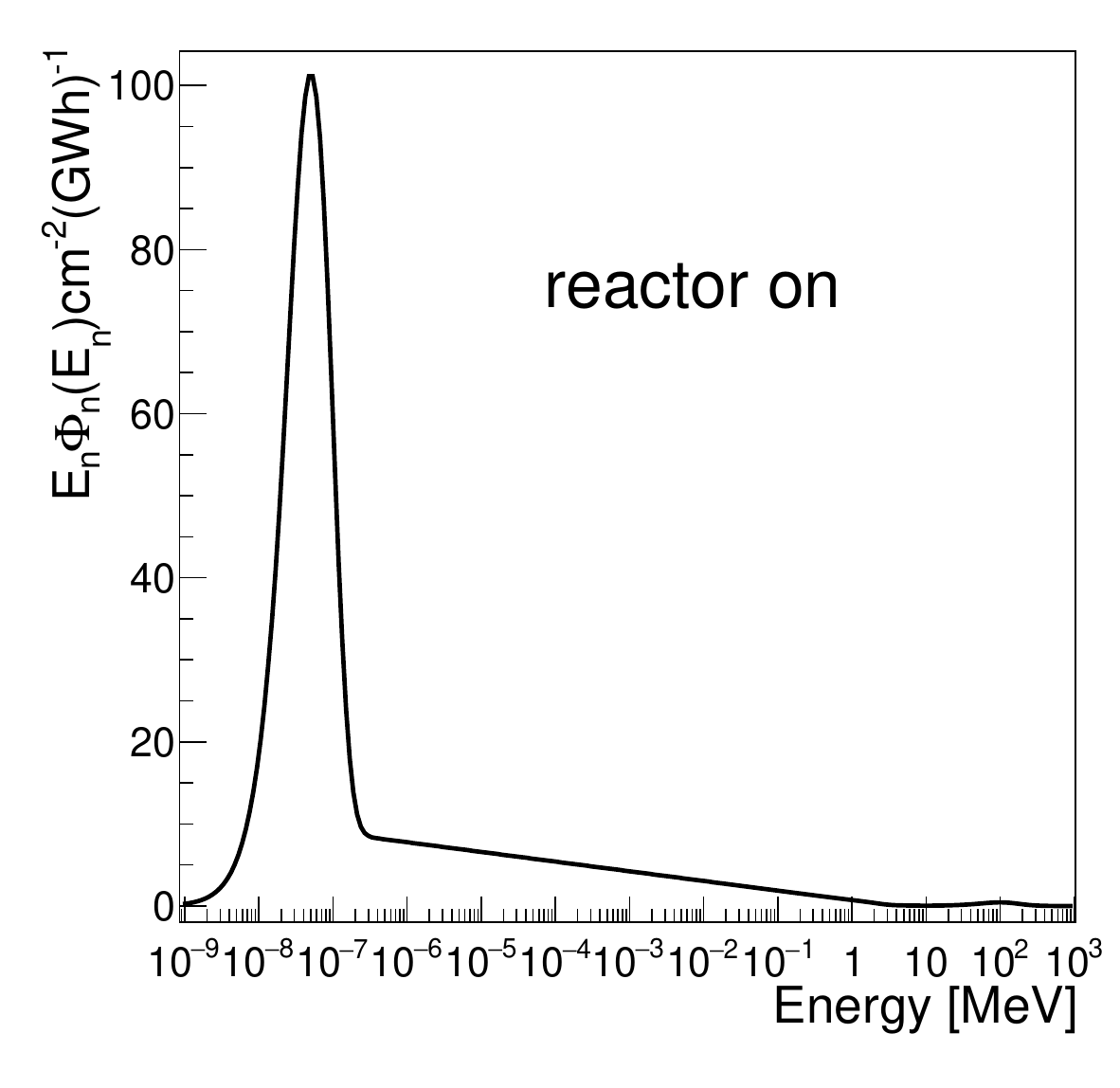}
    \caption{Measured neutron energy distribution $\phi_{on}$(E$_{n})$ resulting from the analysis of the reactor on data normalized to the energy emitted by the reactor.}
    \label{spectra_ON}
\end{figure}

The calculated neutron energy distribution $\phi_{on}$(E$_{n})$  is represented in figure~\ref{spectra_ON} with respect to the neutron energy, E$_{n}$, in terms of the lethargy representation. Due to the low count rates and shape of the spectrum, the parameters of the parametrized model described above were reduced by fixing the location of the evaporation and cascade peak as expected from Monte Carlo simulations (see Sec.\,\ref{sec:4.2}).\\  
The measured reactor spectrum shape is similar to the one obtained in~\cite{Hakenmuller:2019ecb}, with a thermal peak in the sub-eV region, a no-flat intermediate region up to a few keV and a bump around 1~MeV produced mainly by muon-induced neutrons in the reactor building. In addition, an tiny peak is also visible at 100~MeV produced by cosmogenic neutrons. The integral quantity of the neutron fluence normalized to the KKL thermal energy is shown in table~\ref{on_flux} for the individual E$_{n}$ regions of the neutron energy distribution. The energy regions are defined as follows: thermal [$1.0\cdot 10^{-9}$, $4.0\cdot 10^{-7}$]~MeV; intermediate [$4.0\cdot 10^{-7}$, 0.1]~MeV; fast [0.1, 20]~MeV; and cascade [20, 1000]~MeV. Due to the lack of resolution at high-energies, the fast and cascade regions are integrated together. Similar to the KBR, the neutron field is highly thermalized with about 75\% of the neutrons with energies below 0.4~eV. The overall neutron fluence is a factor 30 stronger than in CONUS at KBR. This much larger neutron contribution was already anticipated from the indirect estimations using the $^{63}$Cu $\gamma$-lines from the CONRAD cryostat (Sec.\,\ref{sec:2p1}). However, simulations show a negligible impact over the \conusplus~background, with an impact in the [0.15, 1.0]~keV$_{ee}$ energy region, more than one order of magnitude below the expected \CEvNS~signal.  

\begin{table}[bht]
\caption{Integrated neutron fluence from the reactor-on measurement and normalized to the KKL thermal energy output of 1~GW h. The energy regions are defined as follows: thermal [$1.0\cdot 10^{-9}$, $4.0\cdot 10^{-7}$]~MeV; intermediate [$4.0\cdot 10^{-7}$, 0.1]~MeV; fast + cascade [0.1, 1000]~MeV. }
\label{on_flux}
\centering
\setlength\extrarowheight{4pt}
\begin{tabular}{lc}
\hline
Energy region & $\phi$ [cm$^{-2}$ (GW h)$^{-1}$]\Tstrut\Bstrut\\
\hline
thermal & 195.7$\pm$18.6 \Tstrut\Bstrut\\
intermediate & 62.4$\pm$4.3 \Tstrut\Bstrut\\
fast + cascade & 4.2$\pm$3.4 \Tstrut\Bstrut\\
\hline
total & 262.3$\pm$13.1 \Tstrut\Bstrut\\
\hline
\end{tabular}
\end{table} 

\subsection{Neutrons during reactor off time}
\label{sec:4.2}

The data set during reactor off was acquired with 5 SP9 counters at the same time due to the limited available time. The set used during reactor off period consisted of one bare counter, three counters embedded in the 3.5\inches, 8\inches and 12\inches~polyethylene spheres, and another counter using the 8\inches~sphere with the Cu shell. The spheres were placed around position 2 in figure~\ref{conusplus_room}, with a separation between spheres of only 30~cm due to space constraints. A total measurement time of 5~days was acquired in this configuration. As expected, during reactor off time the count rate decreased several orders of magnitude, but still remains higher than in KBR during reactor off periods. This is consistent with the higher cosmic-ray neutron contribution expected from the smaller overburden. However, the short measurement time and the close distance between the different spheres limited the possibility of obtaining a reliable spectrum with the unfolding method previously employed.

For this reason, an alternative approach was followed based on simulations as done in other experiments~\cite{Goupy:2024zfq,Ricochet:2022pzj}. The cosmogenic neutrons were isotropically generated following the experimental spectrum measured on surface in~\cite{Goldhagen:2004} and propagated through the well-known reactor building geometry until the ZA28R027 room. The shape of the outdoor ground-level neutron spectrum above 5~MeV does not change significantly with the altitude, the vertical geomagnetic cutoff rigidity, R$_{c}$, or the solar modulation~\cite{Gordon:2004}, becoming only necessary to correct for them in the total neutron flux normalization.  A total neutron flux of $(1.4\pm0.2)\cdot 10^{-2}$~neutrons~s$^{-1}$cm$^{-2}$ on surface is estimated using~\cite{Gordon:2004} at the  KKL location, considering a  R$_{c}$ of $\sim$4~GeV and an elevation of 366~m. A similar value was found in~\cite{Goupy:2024zfq}. The second contributor to the neutron spectra during reactor off are the muon-induced neutrons produced in the reactor building, in particular in the concrete of the  ZA28R027 room. For this purpose, muons are generated isotropically following the spectrum from~\cite{Bogdanova:2006ex} over the \conusplus~room with a normalization factor able to directly reproduce the flux measured in Sec.\,\ref{sec:3}.    

\begin{figure}
    \centering
    \includegraphics[width=0.48\textwidth]{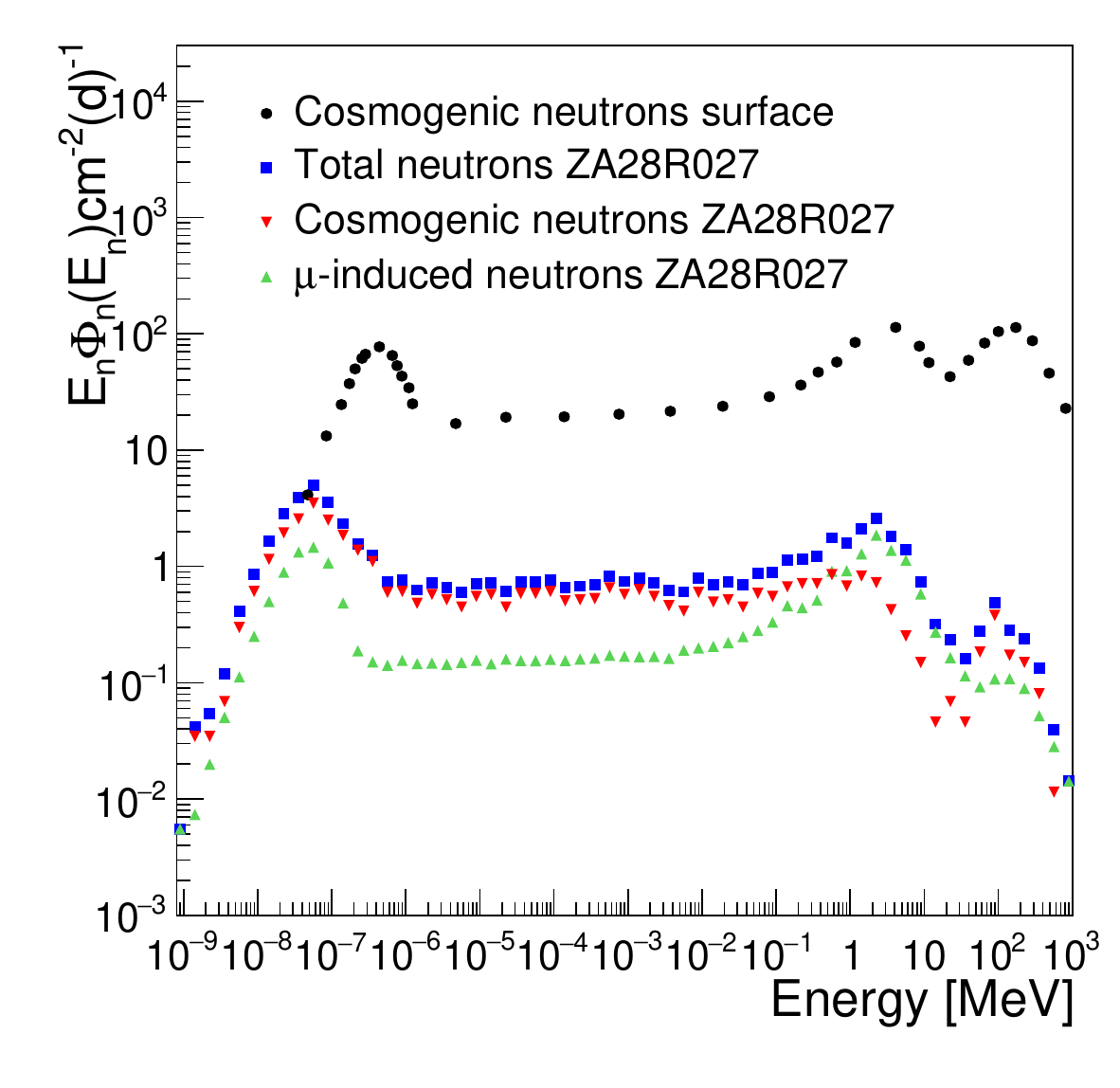}
    \caption{Simulated neutron spectra during reactor off in the  (blue) ZA28R027 room. The muon-induced neutrons generated in the reactor building (green) produce a peak around 1~MeV. The cosmogenic neutrons (red) are reduced by almost 2 orders of magnitude with respect to the surface (black). A peak around 100~MeV is produced by the latter component.}
    \label{spectra_off}
\end{figure}

In this way, $4\cdot 10^{8}$ primary neutrons and $4\cdot 10^{8}$ primary muons have been generated, which corresponds to an equivalent acquisition time of 28~days and 36~days, respectively. The simulations were performed with the framework MaGe~\cite{Boswell:2010mr}, based on Geant4.10.3p03 (\cite{GEANT4:2002zbu, Allison:2006ve}). MaGe is well validated for low energetic electromagnetic processes and the neutron creation and propagation through materials as demonstrated in~\cite{Bonet:2021wjw}. In figure~\ref{spectra_off} the calculated neutron spectra during reactor off is depicted (blue). Two components are depicted: cosmogenic neutrons (red) and muon-induced neutrons from the reactor building (green). The original cosmogenic neutron flux on surface outside the reactor building is also included as reference (black). The total cosmogenic neutron fluence is suppressed almost 2 orders of magnitude, but still a no negligible component over 20~MeV is observed, which will be refereed as "cascade neutrons" in the following. The impact of the reactor drywell head was also studied, observing a 16\% neutron fluence reduction when it is placed over the ZA28R027 room. This effect will be properly accounted in the future \conusplus~background model. The spectra around 1~MeV is dominated by muon-induced neutrons.

The integral quantity of the neutron fluence per day is shown in table~\ref{on_flux} for the individual E$_{n}$ regions of the neutron energy distribution. The energy regions are defined as follows: thermal [$1.0\cdot 10^{-9}$, $4.0\cdot 10^{-7}$]~MeV; intermediate [$4.0\cdot 10^{-7}$, 0.1]~MeV; fast [0.1, 20]~MeV; and cascade [20, 1000]~MeV. A total neutron flux of (28.2$\pm$5.3)~neutrons~d$^{-1}$cm$^{-2}$ is calculated, a factor 2~larger than in KBR, as expected due to the smaller overburden. The neutron component below 20~MeV has an impact below 4\% of the total background rate in the [0.4, 1.0]~keV$_{ee}$ energy region. On the other hand, the cascade neutrons have a large impact for the \conusplus~background, accounting for more than 50\% of the background rate in the  [0.4, 1.0]~keV$_{ee}$ energy region after selection cuts. Future direct measurements are planned to reduce the uncertainty of this background component and mitigation strategies are under discussion. 

\begin{table}[bht]
\caption{Integrated neutron fluence from the reactor-off simulation and normalized to 1 day. The energy regions are defined as follows: thermal [$1.0\cdot 10^{-9}$, $4.0\cdot 10^{-7}$]~MeV; intermediate [$4.0\cdot 10^{-7}$, 0.1]~MeV; fast [0.1, 20]~MeV; cascade [20, 1000]~MeV }
\label{off_flux}
\centering
\setlength\extrarowheight{4pt}
\begin{tabular}{lc}
\hline
Energy region & $\phi$ [cm$^{-2}$ d$^{-1}$]\Tstrut\Bstrut\\
\hline
thermal & 11.5$\pm$2.2 \Tstrut\Bstrut\\
intermediate & 8.1$\pm$1.6 \Tstrut\Bstrut\\
fast & 7.6$\pm$1.1 \Tstrut\Bstrut\\
cascade & 0.9$\pm$0.2 \Tstrut\Bstrut\\
\hline
total & 28.1$\pm$5.3 \Tstrut\Bstrut\\
\hline
\end{tabular}
\end{table} 

In conclusion, the neutron flux was thoroughly studied for the \conusplus~experiment in the new location at the KKL power plant, using direct measurements and simulations. The reactor neutron component is negligible for the \CEvNS~detection, but a much larger cosmogenic neutron component than in CONUS is observed due to the reduced overburden (only 7.4~m~w.e.).

\section{Summary and conclusions}
\label{conclusions}

Neutrino experiments located in closest vicinity to artificial neutrino sources such as commercial nuclear reactor cores profit from the immense neutrino fluxes, however they have to cope with relatively high background levels due to the cosmic radiation present at shallow depth and the artificial radioactivity in such locations. Thus, a deep understanding of all background sources is of fundamental importance. It is needed to optimize the shield design of such neutrino detectors. Further it allows to simulate the residual background seen by the neutrino detectors inside their shield. Finally it allows to clarify, whether problematic background components persist, which are correlated with the reactor thermal power generation and which can potentially mimic the neutrino signal.\\

For the \conusplus~experiment at the Leibstadt nuclear power plant (KKL) a detailed background characterization campaign inside the assigned experimental room was performed involving different techniques: neutron spectroscopy with Bonner spheres, gamma-ray spectroscopy with germanium detectors, muon flux measurement with liquid scintillators, airborne radon detection with a radon sensitive device, and surface contamination via alpha-beta counters and low background germanium spectroscopy. The applied measurement protocol and instrumentation are similar to the ones used for the former CONUS experiment, which was carried out at the Brokdorf nuclear power plant (KBR). This allows for a direct comparison of the main characteristics of the two locations, as summarized in table~\ref{comparison_KBR_KKL} and discussed in the following.\\

\begin{table*}
\begin{center}
\begin{footnotesize}
\caption{Comparison of the CONUS and \conusplus~experimental locations in terms of reactor and environmental conditions as well as background contributions relevant for neutrino searches. Information regarding CONUS at KBR is taken from \cite{Bonet:2020ntx, Hakenmuller:2019ecb, Bonet:2021wjw}. Results about the gamma-radiation refer to reactor on periods only, while the neutron fluence rates are reported for reactor on and off (r. on/off) periods at both locations. The fast neutron fluence during reactor on periods at KKL is calculated subtracting the cascade neutron flux estimated during reactor off periods.
\label{comparison_KBR_KKL} }
\centering
\begin{tabular}{lllc}
\hline
Parameter &  CONUS at KBR &  \conusplus~at KKL \Tstrut\Bstrut\\
\hline
Reactor:            &                                       &                            \Tstrut\Bstrut\\  
\hspace{0.2 cm}  type:             &PWR (3rd gen., pre-Konvoi)     &BWR/6 Mark-III        \Tstrut\Bstrut\\
\hspace{0.2 cm}  make and model:   &KWU (Siemens \& AEG)               &General Electric      \Tstrut\Bstrut\\
\hspace{0.2 cm}  operational:      &1986-2021               &1984-today      \Tstrut\Bstrut\\
\hspace{0.2 cm}  fuel rods:      &193$\times$ UO$_2$/MOX               &648$\times$ UO$_2$     \Tstrut\Bstrut\\
\hspace{0.2 cm}  maximum thermal power: & 3.9~GW        & 3.6~GW \Tstrut\Bstrut\\
\hspace{0.2 cm}  electric gross power:    & 1.480~GW         & 1.285~GW \Tstrut\Bstrut\\
\hline
Experimental location:          &                                       &                           & \Tstrut\Bstrut\\ 
\hspace{0.2 cm} distance to reactor core:    & 17.1~m &  20.7~m\Tstrut\Bstrut\\
\hspace{0.2 cm} neutrino flux:              & 2.3$\cdot$10$^{13}$ s$^{-1}$cm$^{-2}$ & 1.4$\cdot$10$^{13}$ s$^{-1}$cm$^{-2}$ \Tstrut\Bstrut\\
\hspace{0.2 cm} average temperature:       & (26$\pm$3)$^{\circ}$C & (29$\pm$4)$^{\circ}$C \Tstrut\Bstrut\\
\hspace{0.2 cm} averaged overburden:       & 24~m w.e. & 7.4~m w.e.\Tstrut\Bstrut\\
\hspace{0.2 cm} height above sea level:    & 11~m a.s.l. & 366~m a.s.l.\Tstrut\Bstrut\\
\hspace{0.2 cm} geographical coordinates:  & 53.850833$^{\circ}$N, 9.344722$^{\circ}$E & 47.60135$^{\circ}$N , 8.18375$^{\circ}$E\Tstrut\Bstrut\\
\hline
Background radiation:          &                                       &                            \Tstrut\Bstrut\\ 
\hspace{0.2 cm} cosmic muon flux: & (47$\pm$3)~muons~s$^{-1}$m$^{-2}$ &  (107$\pm$3)~muons~s$^{-1}$m$^{-2}$\Tstrut\Bstrut\\
\hspace{0.2 cm} radon concentration in air:  & (175$\pm$35) Bq m$^{-3}$  & (110$\pm$80) Bq m$^{-3}$ \Tstrut\Bstrut\\
\hspace{0.2 cm} $\gamma$-ray rate:  &   &     \Tstrut\Bstrut\\
\hspace{0.5 cm} primary isotopes:  &$^{16}$N           &$^{56}$Fe, $^{28}$Si, $^{40}$Ca                \Tstrut\Bstrut\\
\hspace{0.5 cm} high energy ($\geq$ 2.7~MeV$_{ee}$)  & 155~counts~kg$^{-1}$ s$^{-1}$  & 6~counts~kg$^{-1}$ s$^{-1}$\Tstrut\Bstrut\\
\hspace{0.5 cm} low energy ($\leq$ 2.7~MeV$_{ee}$)  & 690~counts~kg$^{-1}$ s$^{-1}$ & 670~counts~kg$^{-1}$s$^{-1}$ \Tstrut\Bstrut\\ 
\hspace{0.2 cm} surface contamination:&                     &                           & \Tstrut\Bstrut\\
\hspace{0.5 cm} primary isotopes:  &$^{60}$Co, $^{137}$Cs           &$^{60}$Co, $^{54}$Mn                          & \Tstrut\Bstrut\\
\hspace{0.5 cm} secondary isotopes:&$^{134}$Cs, $^{54}$Mn, $^{m110}$Ag, $^{125}$Sb   &  $^{156}$Eu	                      \Tstrut\Bstrut\\
\hspace{0.2 cm} neutron fluence rate: &  & & \Tstrut\Bstrut\\ 
\hspace{0.5 cm} thermal (r. on)& (601$\pm$38) cm$^{-2}$d$^{-1}$ & (16908$\pm$1602) cm$^{-2}$d$^{-1}$  \Tstrut\Bstrut\\ 
\hspace{0.5 cm} thermal (r. off)& (4.5$\pm$0.7) cm$^{-2}$d$^{-1}$& (11.5$\pm$2.2) cm$^{-2}$d$^{-1}$ \Tstrut\Bstrut\\ 
\hspace{0.5 cm} intermediate (r. on)& (146$\pm$20) cm$^{-2}$d$^{-1}$ & (5391$\pm$368) cm$^{-2}$d$^{-1}$  \Tstrut\Bstrut\\ 
\hspace{0.5 cm} intermediate (r. off)& (4.2$\pm$1.2) cm$^{-2}$d$^{-1}$& (8.1$\pm$1.6) cm$^{-2}$d$^{-1}$  \Tstrut\Bstrut\\ 
\hspace{0.5 cm} fast (r. on)& (14.1$\pm$4.7) cm$^{-2}$d$^{-1}$ & (362$\pm$290) cm$^{-2}$d$^{-1}$  \Tstrut\Bstrut\\ 
\hspace{0.5 cm} fast (r. off)& (6.4$\pm$1.0) cm$^{-2}$d$^{-1}$ & (7.6$\pm$1.1) cm$^{-2}$d$^{-1}$  \Tstrut\Bstrut\\ 
\hspace{0.5 cm} cosmic neutrons & negligible & (0.9$\pm$0.2) cm$^{-2}$d$^{-1}$  \Tstrut\Bstrut\\ 
\hline
\end{tabular}
\end{footnotesize}
\end{center}
\end{table*} 

The overburden as well the residual cosmic muons at the experimental locations were measured with a liquid scintillator detector. For KKL, an averaged overburden of 7.4~m~w.e. was determined. This is by factor 3 smaller than the one achieved at KBR. The overburden at KKL is large enough to suppress the hadronic component by almost 2 orders of magnitude. However, simulations show a no negligible contribution from cosmogenic related neutrons over 20~MeV with a flux of almost 1~neutron~d$^{-1}$cm$^{-2}$, becoming the main background component in the \CEvNS~region of interest. Regarding the muon flux, it is reduced only by factor ~2 compared to surface and needs therefore a better rejection than at KBR. The muon-induced neutrons in the reactor building accounts for 4\% of the \conusplus~background in the [0.4, 1.0]~keV$_{ee}$ energy region. \\

The mean air-borne radon concentration was found to be in both locations similarly around $110-175$~Bq~m$^{-3}$. Values of a few 100~Bq~m$^{-3}$ seem to be normal in such locations, considering the fact that the experimental locations are within the safety containments of a reactor, in which fresh air exchange is limited, and are surrounded by massive structural components (concrete walls of biological shields etc.), which naturally emanate radon. In order to mitigate airborne radon, bottles filled with breathing air, which were stored for a few weeks in order to allow the radon to decay away, are used to permanently flush the detector chambers in CONUS and \conusplus~with radon-free air.\\

The gamma-radiation was studied with the same HPGe detector CONRAD up to an energy of  11~MeV$_{ee}$. Similar radioactivity levels were found for the natural radioactivity occurring below 2.7~MeV$_{ee}$. For the artificial $^{60}$Co isotope with its characteristic gamma-rays at 1.1~MeV, 1.3~MeV and~2.5 MeV different values were found and further analyzed in terms of surface vs. bulk contamination, with a 5~times stronger presence in KKL. Regarding the radioactivity above 2.7~MeV$_{ee}$, which is mainly induced by neutron escaping the reactor core and captured by isotopes in the water cooling cycle or in the construction materials, it turned out to be 26 times smaller at the KKL site compared to the KBR site. The reason is, that the water cooling pipe systems, which contain the main gamma-emitter $^{16}$N, is by far more distant from the experimental site at KKL than at KBR.\\

With the help of wipe tests it was possible to distinguish further surface from bulk contamination and to quantify the risk of a potential contamination of the experimental setup during assembly/disassembly and operation. The determined nuclide vectors from wipe tests of room and shield surfaces at the KKL and KBR experimental site turned out to be different, reflecting differences in type, structure, materials of the two reactors. The study confirmed also, that the installation of the shield-detector setup can be performed without any relevant cross-contamination, if proper counteractions are imposed.\\

The neutron fluence was measured with a Bonner Sphere \cite{BRAMBLETT19601,NEMUS2002} in scientific cooperation with the neutron divisions of the following institutions: the Physikalisch-Technische Bundesanstalt for CONUS at KBR, and the Paul Scherrer Institut for \conusplus~at KKL, known for their long-lasting experience in neutron spectroscopy.  The measurements were performed during reactor on and off periods to study the reactor correlation neutron background and the permanent cosmogenic neutron background. However, the measurement during reactor off at KKL was not conclusive and a complementary method with simulations was followed. A total neutron fluence of  $(2.3\pm0.1)\cdot 10^{4}$~neutrons~d$^{-1}$cm$^{-2}$ was measured at KKL during reactor on, a factor 30~times larger than at KBR. For reactor off periods, simulations evidenced a cosmogenic neutron flux of (28.2$\pm$5.3)~neutrons~d$^{-1}$cm$^{-2}$, 2~times larger than in KBR. This is consistent with the smaller overburden measured for \conusplus. \\

The intercomparison of the background conditions for \conusplus~at KKL and for CONUS at KBR led to the following actions and conclusions:\\
First,  the different background levels encountered at KKL and KBR motivated us to optimize the original CONUS shield for \conusplus: we removed one out of five lead layers (20 instead of 25 cm), since the high-energy gamma-radiation turned out to be less pronounced in the case of the KKL experimental site. The extra free space was used to integrate a second muon veto system to enhance the overall muon veto efficiency needed at KKL, where the muon flux is larger due to a smaller overburden. With this counteraction we aim to reach similar muon background contribution in both experiments.\\
Second, the smaller overburden significantly increased the cosmogenic neutron flux in the \conusplus~location. Neutrons over 20~MeV are the dominant component for the background spectra in the \conusplus~detectors below a few keV. Future actions will be taken to mitigate this background.\\   
Third, background components correlated with the reactor thermal power, i.e.  reactor neutrons and neutron-capture induced  isotopes with decay times of less than a few tens of seconds, are highly problematic, since they can mimic neutrino signals. We were able to measure all these components outside the shield-detector-setup and to simulate their impact on the neutrino detectors inside the shield. On this way, we could proof that their impact on the region of interest for both experiments CONUS and \conusplus~is negligibly small.\\
Fourth,  the different background levels encountered at KKL and KBR demonstrates that it is not possible to extrapolate the background composition from one location to the other. Even within one experimental room the background conditions can change within a few meters. Thus, for each exact location, in which later on a neutrino detector together with its shield will be installed, it is mandatory to conduct a dedicated background characterization campaign for a correct interpretation of the neutrino detector data. The methods and techniques applied in this work might serve as an example and guidance for all upcoming reactor neutrino experiments.\\

\small

\textbf{Acknowledgements}

For technical, mechanical, electronics, DAQ, IT and logistical support we thank all involved divisions and workshops at the Max-Planck-Institut f\"ur Kernphysik in Heidelberg. 
In particular we thank T. Frydlewicz in transport and custom issues of the experiment and  M. Reissfelder for the technical support. We thank J. Schreiner and D. Gross from the MPIK radiation protection department from their continuous support.   
We express our deepest gratitude to the Leibstadt AG for hosting and supporting the \conusplus~experiment.
For the support during the measurements performed in this publication we specifically thank to R. Gaalev from PSI and R. Meili, A. Ritter, M. Sailer and all involved departments at KKL.
We thank R. Bieli for providing the Leibstadt reactor thermal power and neutron monitoring sensors data. The \conusplus~experiment is supported financially by the Max Planck Society (MPG).

\normalsize


\bibliographystyle{bibliostyle}
\bibliography{references}

\begin{thebibliography}{10}
\expandafter\ifx\csname urlstyle\endcsname\relax
  \providecommand{\doi}[1]{doi:\discretionary{}{}{}#1}
  \providecommand{\eprint}[1]{arXiv:\discretionary{}{}{}#1}\else
  \providecommand{\doi}{doi:\discretionary{}{}{}\begingroup \urlstyle{rm}\Url}
  \providecommand{\eprint}{arXiv:\discretionary{}{}{}\begingroup
  \urlstyle{rm}\Url}\fi

\bibitem{Freedman:1973yd}
D.~Z. Freedman.
\newblock \emph{{Coherent Neutrino Nucleus Scattering as a Probe of the Weak
  Neutral Current}}.
\newblock Phys. Rev. D, \textbf{9}:1389--1392, 1974.
\newblock
  \href{http://dx.doi.org/10.1103/PhysRevD.9.1389}{\doi{10.1103/PhysRevD.9.1389}}.

\bibitem{Kopeliovich:1974mv}
V.~B. Kopeliovich and L.~L. Frankfurt.
\newblock \emph{{Isotopic and chiral structure of neutral current}}.
\newblock JETP Lett., \textbf{19}:145--147, 1974.

\bibitem{Coherent:2017}
D.~Akimov et~al.
\newblock \emph{{Observation of Coherent Elastic Neutrino-Nucleus Scattering}}.
\newblock Science, \textbf{357}(6356):1123--1126, 2017.
\newblock
  \href{http://dx.doi.org/10.1126/science.aao0990}{\doi{10.1126/science.aao0990}}.
\newblock \href{https://arxiv.org/abs/1708.01294}{\eprint{1708.01294}}.

\bibitem{COHERENT:2020iec}
D.~Akimov et~al.
\newblock \emph{{First Measurement of Coherent Elastic Neutrino-Nucleus
  Scattering on Argon}}.
\newblock Phys. Rev. Lett., \textbf{126}(1):012002, 2021.
\newblock
  \href{http://dx.doi.org/10.1103/PhysRevLett.126.012002}{\doi{10.1103/PhysRevLett.126.012002}}.
\newblock \href{https://arxiv.org/abs/2003.10630}{\eprint{2003.10630}}.

\bibitem{Adamski:2024yqt}
S.~Adamski et~al.
\newblock \emph{{First detection of coherent elastic neutrino-nucleus
  scattering on germanium}}.
\newblock 2024.
\newblock \href{https://arxiv.org/abs/2406.13806}{\eprint{2406.13806}}.

\bibitem{XENON:2024ijk}
E.~Aprile et~al.
\newblock \emph{{First Indication of Solar B8 Neutrinos via Coherent Elastic
  Neutrino-Nucleus Scattering with XENONnT}}.
\newblock Phys. Rev. Lett., \textbf{133}(19):191002, 2024.
\newblock
  \href{http://dx.doi.org/10.1103/PhysRevLett.133.191002}{\doi{10.1103/PhysRevLett.133.191002}}.
\newblock \href{https://arxiv.org/abs/2408.02877}{\eprint{2408.02877}}.

\bibitem{PandaX:2024muv}
Z.~Bo et~al.
\newblock \emph{{First Indication of Solar B8 Neutrinos through Coherent
  Elastic Neutrino-Nucleus Scattering in PandaX-4T}}.
\newblock Phys. Rev. Lett., \textbf{133}(19):191001, 2024.
\newblock
  \href{http://dx.doi.org/10.1103/PhysRevLett.133.191001}{\doi{10.1103/PhysRevLett.133.191001}}.
\newblock \href{https://arxiv.org/abs/2407.10892}{\eprint{2407.10892}}.

\bibitem{Aguilar-Arevalo:2024dln}
A.~Aguilar-Arevalo.
\newblock \emph{{Upgraded CONNIE experiment with Skipper CCDs / CONNIE First
  results with Skipper-CCDs}}.
\newblock PoS, \textbf{TAUP2023}:154, 2024.
\newblock
  \href{http://dx.doi.org/10.22323/1.441.0154}{\doi{10.22323/1.441.0154}}.

\bibitem{Ackermann:2024kxo}
N.~Ackermann et~al.
\newblock \emph{{Final CONUS results on coherent elastic neutrino nucleus
  scattering at the Brokdorf reactor}}.
\newblock Phys. Rev. Lett., \textbf{133}:251802, 2024.
\newblock
  \href{http://dx.doi.org/10.1103/PhysRevLett.133.251802}{\doi{10.1103/PhysRevLett.133.251802}}.

\bibitem{MINER:2016igy}
G.~Agnolet et~al.
\newblock \emph{{Background Studies for the MINER Coherent Neutrino Scattering
  Reactor Experiment}}.
\newblock Nucl. Instrum. Meth. A, \textbf{853}:53--60, 2017.
\newblock
  \href{http://dx.doi.org/10.1016/j.nima.2017.02.024}{\doi{10.1016/j.nima.2017.02.024}}.
\newblock \href{https://arxiv.org/abs/1609.02066}{\eprint{1609.02066}}.

\bibitem{Colaresi:2022obx}
J.~Colaresi, J.~I. Collar, T.~W. Hossbach, C.~M. Lewis and K.~M. Yocum.
\newblock \emph{{Measurement of Coherent Elastic Neutrino-Nucleus Scattering
  from Reactor Antineutrinos}}.
\newblock Phys. Rev. Lett., \textbf{129}(21):211802, 2022.
\newblock
  \href{http://dx.doi.org/10.1103/PhysRevLett.129.211802}{\doi{10.1103/PhysRevLett.129.211802}}.
\newblock \href{https://arxiv.org/abs/2202.09672}{\eprint{2202.09672}}.

\bibitem{NEON:2022hbk}
J.~J. Choi et~al.
\newblock \emph{{Exploring coherent elastic neutrino-nucleus scattering using
  reactor electron antineutrinos in the NEON experiment}}.
\newblock Eur. Phys. J. C, \textbf{83}(3):226, 2023.
\newblock
  \href{http://dx.doi.org/10.1140/epjc/s10052-023-11352-x}{\doi{10.1140/epjc/s10052-023-11352-x}}.
\newblock \href{https://arxiv.org/abs/2204.06318}{\eprint{2204.06318}}.

\bibitem{NUCLEUS:2022zti}
C.~Goupy et~al.
\newblock \emph{{Exploring coherent elastic neutrino-nucleus scattering of
  reactor neutrinos with the NUCLEUS experiment}}.
\newblock SciPost Phys. Proc., \textbf{12}:053, 2023.
\newblock
  \href{http://dx.doi.org/10.21468/SciPostPhysProc.12.053}{\doi{10.21468/SciPostPhysProc.12.053}}.
\newblock \href{https://arxiv.org/abs/2211.04189}{\eprint{2211.04189}}.

\bibitem{nGeN:2022uje}
I.~Alekseev et~al.
\newblock \emph{{First results of the \ensuremath{\nu}GeN experiment on
  coherent elastic neutrino-nucleus scattering}}.
\newblock Phys. Rev. D, \textbf{106}(5):L051101, 2022.
\newblock
  \href{http://dx.doi.org/10.1103/PhysRevD.106.L051101}{\doi{10.1103/PhysRevD.106.L051101}}.
\newblock \href{https://arxiv.org/abs/2205.04305}{\eprint{2205.04305}}.

\bibitem{Akimov:2024lnl}
D.~Y. Akimov et~al.
\newblock \emph{{First constraints on the coherent elastic scattering of
  reactor antineutrinos off xenon nuclei}}.
\newblock 2024.
\newblock \href{https://arxiv.org/abs/2411.18641}{\eprint{2411.18641}}.

\bibitem{Yang:2024exl}
L.~Yang, Y.~Liang and Q.~Yue.
\newblock \emph{{RECODE program for reactor neutrino CEvNS detection with PPC
  Germanium detector}}.
\newblock PoS, \textbf{TAUP2023}:296, 2024.
\newblock
  \href{http://dx.doi.org/10.22323/1.441.0296}{\doi{10.22323/1.441.0296}}.

\bibitem{RELICS:2024opj}
C.~Cai et~al.
\newblock \emph{{Reactor neutrino liquid xenon coherent elastic scattering
  experiment}}.
\newblock Phys. Rev. D, \textbf{110}(7):072011, 2024.
\newblock
  \href{http://dx.doi.org/10.1103/PhysRevD.110.072011}{\doi{10.1103/PhysRevD.110.072011}}.
\newblock \href{https://arxiv.org/abs/2405.05554}{\eprint{2405.05554}}.

\bibitem{Ricochet:2023nvt}
C.~Augier et~al.
\newblock \emph{{First demonstration of 30 eVee ionization energy resolution
  with Ricochet germanium cryogenic bolometers}}.
\newblock Eur. Phys. J. C, \textbf{84}(2):186, 2024.
\newblock
  \href{http://dx.doi.org/10.1140/epjc/s10052-024-12433-1}{\doi{10.1140/epjc/s10052-024-12433-1}}.
\newblock \href{https://arxiv.org/abs/2306.00166}{\eprint{2306.00166}}.

\bibitem{TEXONO:2024vfk}
S.~Karmakar et~al.
\newblock \emph{{New Limits on Coherent Neutrino Nucleus Elastic Scattering
  Cross Section at the Kuo-Sheng Reactor Neutrino Laboratory}}.
\newblock 2024.
\newblock \href{https://arxiv.org/abs/2411.18812}{\eprint{2411.18812}}.

\bibitem{CONUS:2024lnu}
N.~Ackermann et~al.
\newblock \emph{{CONUS+~Experiment}}.
\newblock Eur. Phys. J. C, \textbf{84}(12):1265, 2024.
\newblock
  \href{http://dx.doi.org/10.1140/epjc/s10052-024-13551-6}{\doi{10.1140/epjc/s10052-024-13551-6}}.
\newblock \href{https://arxiv.org/abs/2407.11912}{\eprint{2407.11912}}.

\bibitem{THOMAS200212}
D.~Thomas and A.~Alevra.
\newblock \emph{Bonner sphere spectrometers—a critical review}.
\newblock Nucl. Instrum. Methods Phys. Res., \textbf{476}(1):12--20, 2002.
\newblock ISSN 0168-9002.
\newblock
  \href{http://dx.doi.org/https://doi.org/10.1016/S0168-9002(01)01379-1}{\doi{https://doi.org/10.1016/S0168-9002(01)01379-1}},
  int. Workshop on Neutron Field Spectrometry in Science, Technolog y and
  Radiation Protection.

\bibitem{Hakenmuller:2019ecb}
J.~Hakenm\"uller et~al.
\newblock \emph{{Neutron-induced background in the CONUS experiment}}.
\newblock Eur. Phys. J. C, \textbf{79}(8):699, 2019.
\newblock
  \href{http://dx.doi.org/10.1140/epjc/s10052-019-7160-2}{\doi{10.1140/epjc/s10052-019-7160-2}}.
\newblock \href{https://arxiv.org/abs/1903.09269}{\eprint{1903.09269}}.

\bibitem{Bogdanova:2006ex}
L.~N. Bogdanova, M.~G. Gavrilov, V.~N. Kornoukhov and A.~S. Starostin.
\newblock \emph{{Cosmic muon flux at shallow depths underground}}.
\newblock Phys. Atom. Nucl., \textbf{69}:1293--1298, 2006.
\newblock
  \href{http://dx.doi.org/10.1134/S1063778806080047}{\doi{10.1134/S1063778806080047}}.
\newblock
  \href{https://arxiv.org/abs/nucl-ex/0601019}{\eprint{nucl-ex/0601019}}.

\bibitem{ParticleDataGroup:2020ssz}
P.~A. Zyla et~al.
\newblock \emph{{Review of Particle Physics}}.
\newblock PTEP, \textbf{2020}(8):083C01, 2020.
\newblock
  \href{http://dx.doi.org/10.1093/ptep/ptaa104}{\doi{10.1093/ptep/ptaa104}}.

\bibitem{Bonet:2021wjw}
H.~Bonet et~al.
\newblock \emph{{Full background decomposition of the CONUS experiment}}.
\newblock Eur. Phys. J. C, \textbf{83}(3):195, 2023.
\newblock
  \href{http://dx.doi.org/10.1140/epjc/s10052-023-11240-4}{\doi{10.1140/epjc/s10052-023-11240-4}}.
\newblock \href{https://arxiv.org/abs/2112.09585}{\eprint{2112.09585}}.

\bibitem{Bonet:2020ntx}
H.~Bonet et~al.
\newblock \emph{{Large-size sub-keV sensitive germanium detectors for the CONUS
  experiment}}.
\newblock Eur. Phys. J. C, \textbf{81}(3):267, 2021.
\newblock
  \href{http://dx.doi.org/10.1140/epjc/s10052-021-09038-3}{\doi{10.1140/epjc/s10052-021-09038-3}}.
\newblock \href{https://arxiv.org/abs/2010.11241}{\eprint{2010.11241}}.

\bibitem{Hakenmuller:2023yzb}
J.~Hakenm\"uller and G.~Heusser.
\newblock \emph{{CONRAD\textemdash{}A low level germanium test detector for the
  CONUS experiment}}.
\newblock Appl. Radiat. Isot., \textbf{194}:110669, 2023.
\newblock
  \href{http://dx.doi.org/10.1016/j.apradiso.2023.110669}{\doi{10.1016/j.apradiso.2023.110669}}.

\bibitem{Bediako:2015}
M.~Bediako and E.~Amankwah.
\newblock \emph{Analysis of Chemical Composition of Portland Cement in Ghana: A
  Key to Understand the Behavior of Cement}.
\newblock Advances in Materials Science and Engineering, \textbf{Volume 2015}:5
  pages, 2015.
\newblock
  \href{http://dx.doi.org/10.1155/2015/349401}{\doi{10.1155/2015/349401}}.

\bibitem{IAEE:2007}
Database of Prompt Gamma Rays from Slow Neutron Capture for Elemental Analysis.
\newblock Non-serial Publications. IAEA, Vienna, 2007.
\newblock ISBN 92-0-101306-X.

\bibitem{BUDJAS2009706}
D.~Budjáš, M.~Heisel, W.~Maneschg and H.~Simgen.
\newblock \emph{Optimisation of the MC-model of a p-type Ge-spectrometer for
  the purpose of efficiency determination}.
\newblock Appl. Radiat. Isot., \textbf{67}(5):706--710, 2009.
\newblock ISSN 0969-8043.
\newblock
  \href{http://dx.doi.org/https://doi.org/10.1016/j.apradiso.2009.01.015}{\doi{https://doi.org/10.1016/j.apradiso.2009.01.015}}.

\bibitem{RadiologicalInventory98}
IAEA.
\newblock \emph{Radiological characterization of shut down nuclear reactors for
  decommissioning purposes}.
\newblock 1998.
\newblock ISSN 0074-1914, ISBN 92-0-103198-X.

\bibitem{DELMORE20111008}
J.~E. Delmore, D.~C. Snyder, T.~Tranter and N.~R. Mann.
\newblock \emph{Cesium isotope ratios as indicators of nuclear power plant
  operations}.
\newblock Journal of Environmental Radioactivity, \textbf{102}(11):1008--1011,
  2011.
\newblock ISSN 0265-931X.
\newblock
  \href{http://dx.doi.org/https://doi.org/10.1016/j.jenvrad.2011.06.013}{\doi{https://doi.org/10.1016/j.jenvrad.2011.06.013}}.

\bibitem{Bonhomme:2022xcg}
A.~Bonhomme, C.~Buck, B.~Gramlich and M.~Raab.
\newblock \emph{{Safe liquid scintillators for large scale detectors}}.
\newblock JINST, \textbf{17}(11):P11025, 2022.
\newblock
  \href{http://dx.doi.org/10.1088/1748-0221/17/11/P11025}{\doi{10.1088/1748-0221/17/11/P11025}}.
\newblock \href{https://arxiv.org/abs/2205.15046}{\eprint{2205.15046}}.

\bibitem{Hua-Lin:2009gyx}
X.~Hua-Lin.
\newblock \emph{{Oxygen quenching in LAB based liquid scintillator and nitrogen
  bubbling}}.
\newblock Chin. Phys. C, \textbf{34}:571--575, 2010.
\newblock
  \href{http://dx.doi.org/10.1088/1674-1137/34/5/011}{\doi{10.1088/1674-1137/34/5/011}}.
\newblock \href{https://arxiv.org/abs/0904.1329}{\eprint{0904.1329}}.

\bibitem{HAUSSER1993223}
G.~Heusser.
\newblock \emph{{Cosmic ray-induced background in Ge-spectrometry}}.
\newblock Nucl. Instrum. Methods Phys. Res., \textbf{83}(1):223--228, 1993.
\newblock ISSN 0168-583X.
\newblock
  \href{http://dx.doi.org/https://doi.org/10.1016/0168-583X(93)95931-T}{\doi{https://doi.org/10.1016/0168-583X(93)95931-T}}.

\bibitem{NEMUS2002}
B.~Wiegel and A.~Alevra.
\newblock \emph{NEMUS—the PTB Neutron Multisphere Spectrometer: Bonner
  spheres and more}.
\newblock Nuclear Instruments and Methods in Physics Research Section A:
  Accelerators, Spectrometers, Detectors and Associated Equipment,
  \textbf{476}(1):36--41, 2002.
\newblock
  \href{http://dx.doi.org/https://doi.org/10.1016/S0168-9002(01)01385-7}{\doi{https://doi.org/10.1016/S0168-9002(01)01385-7}}.

\bibitem{GALEEV2021}
R.~Galeev.
\newblock Characterization of neutron stray fields for residual activation
  estimation around high-energy accelerators using spectrometric and Monte
  Carlo methods.
\newblock Thesis, University of Basel, Faculty of Science, 2021.

\bibitem{PIAF2009}
S.~Röttger et~al.
\newblock \emph{The PTB neutron reference fields (PIAF)—quasi‐monoenergetic
  neutron reference fields in the energy range from thermal to 200 MeV}.
\newblock AIP Conf. Proc., \textbf{1175}:375–381, 2009.
\newblock
  \href{http://dx.doi.org/https://doi.org/10.1063/1.3258255}{\doi{https://doi.org/10.1063/1.3258255}}.

\bibitem{WIEGEL200252}
B.~Wiegel, A.~Alevra, M.~Matzke, U.~Schrewe and J.~Wittstock.
\newblock \emph{Spectrometry using the PTB neutron multisphere spectrometer
  (NEMUS) at flight altitudes and at ground level}.
\newblock Nucl. Instrum. Methods Phys. Res., \textbf{476}(1):52--57, 2002.
\newblock ISSN 0168-9002.
\newblock
  \href{http://dx.doi.org/https://doi.org/10.1016/S0168-9002(01)01387-0}{\doi{https://doi.org/10.1016/S0168-9002(01)01387-0}}.

\bibitem{REGINATTO2010}
M.~Reginatto.
\newblock \emph{Overview of spectral unfolding techniques and uncertainty
  estimation}.
\newblock Radiation Measurements, \textbf{45}(10):1323--1329, 2010.
\newblock ISSN 1350-4487.
\newblock
  \href{http://dx.doi.org/https://doi.org/10.1016/j.radmeas.2010.06.016}{\doi{https://doi.org/10.1016/j.radmeas.2010.06.016}}.

\bibitem{REGINATTO2009}
M.~Reginatto, E.~, Hohmann and B.~Wiegel.
\newblock \emph{How Accurately Can We Determine Spectra in High-Energy Neutron
  Fields with Bonner Spheres?}
\newblock Nucl. Tech., \textbf{168}:328–332, 2009.

\bibitem{REGINATTO2006}
M.~Reginatto.
\newblock \emph{Bayesian approach for quantifying the uncertainty of neutron
  doses derived from spectrometric measurements}.
\newblock Radiat. Prot. Dosimetry, \textbf{168}:64–69, 2006.
\newblock
  \href{http://dx.doi.org/https://10.1093/rpd/ncl096}{\doi{https://10.1093/rpd/ncl096}}.

\bibitem{JAGS2003}
M.~Plummer.
\newblock \emph{JAGS: A Program for Analysis of Bayesian Graphical Models Using
  Gibbs Sampling}.
\newblock Proceedings of the 3rd International Workshop on Distributed
  Statistical Computing, 2003.

\bibitem{TRS318}
{International Atomic Energy Agency}.
\newblock Compendium of Neutron Spectra and Detector Responses for Radiation
  Protection Purposes.
\newblock Supplement to Technical Reports Series No. 318, Vienna, 2001.

\bibitem{Goupy:2024zfq}
C.~Goupy, S.~Marnieros, B.~Mauri, C.~Nones and M.~Vivier.
\newblock \emph{{Prototyping a High Purity Germanium cryogenic veto system for
  a bolometric detection experiment}}.
\newblock Nucl. Instrum. Meth. A, \textbf{1064}:169383, 2024.
\newblock
  \href{http://dx.doi.org/10.1016/j.nima.2024.169383}{\doi{10.1016/j.nima.2024.169383}}.
\newblock \href{https://arxiv.org/abs/2401.09837}{\eprint{2401.09837}}.

\bibitem{Ricochet:2022pzj}
C.~Augier et~al.
\newblock \emph{{Fast neutron background characterization of the future
  Ricochet experiment at the ILL research nuclear reactor}}.
\newblock Eur. Phys. J. C, \textbf{83}(1):20, 2023.
\newblock
  \href{http://dx.doi.org/10.1140/epjc/s10052-022-11150-x}{\doi{10.1140/epjc/s10052-022-11150-x}}.
\newblock \href{https://arxiv.org/abs/2208.01760}{\eprint{2208.01760}}.

\bibitem{Goldhagen:2004}
P.~Goldhagen, J.~M. Clem and J.~W. Wilson.
\newblock \emph{{The energy spectrum of cosmic-ray induced neutrons measured on
  an airplane over a wide range of altitude and latitude}}.
\newblock Radiation Protection Dosimetry, \textbf{110}(1-4):387--392, 2004.
\newblock ISSN 0144-8420.
\newblock
  \href{http://dx.doi.org/10.1093/rpd/nch216}{\doi{10.1093/rpd/nch216}}.

\bibitem{Gordon:2004}
M.~Gordon et~al.
\newblock \emph{{Measurement of the flux and energy spectrum of cosmic-ray
  induced neutrons on the ground}}.
\newblock IEEE Transactions on Nuclear Science, \textbf{51}(6):3427--3434,
  2004.
\newblock
  \href{http://dx.doi.org/10.1109/TNS.2004.839134}{\doi{10.1109/TNS.2004.839134}}.

\bibitem{Boswell:2010mr}
M.~Boswell et~al.
\newblock \emph{{MaGe-a Geant4-based Monte Carlo Application Framework for
  Low-background Germanium Experiments}}.
\newblock IEEE Trans. Nucl. Sci., \textbf{58}:1212--1220, 2011.
\newblock
  \href{http://dx.doi.org/10.1109/TNS.2011.2144619}{\doi{10.1109/TNS.2011.2144619}}.
\newblock \href{https://arxiv.org/abs/1011.3827}{\eprint{1011.3827}}.

\bibitem{GEANT4:2002zbu}
S.~Agostinelli et~al.
\newblock \emph{{GEANT4--a simulation toolkit}}.
\newblock Nucl. Instrum. Meth. A, \textbf{506}:250--303, 2003.
\newblock
  \href{http://dx.doi.org/10.1016/S0168-9002(03)01368-8}{\doi{10.1016/S0168-9002(03)01368-8}}.

\bibitem{Allison:2006ve}
J.~Allison et~al.
\newblock \emph{{Geant4 developments and applications}}.
\newblock IEEE Trans. Nucl. Sci., \textbf{53}:270, 2006.
\newblock
  \href{http://dx.doi.org/10.1109/TNS.2006.869826}{\doi{10.1109/TNS.2006.869826}}.

\bibitem{BRAMBLETT19601}
R.~L. Bramblett, R.~I. Ewing and T.~Bonner.
\newblock \emph{A new type of neutron spectrometer}.
\newblock Nuclear Instruments and Methods, \textbf{9}(1):1--12, 1960.
\newblock ISSN 0029-554X.
\newblock
  \href{http://dx.doi.org/https://doi.org/10.1016/0029-554X(60)90043-4}{\doi{https://doi.org/10.1016/0029-554X(60)90043-4}}.

\end{thebibliography}

\end{document}